%% file: main.tex
\documentclass[sigconf]{acmart}

\copyrightyear{2024} 
\acmYear{2024} 
\setcopyright{acmlicensed}\acmConference[WSDM '24]{Proceedings of the 17th ACM International Conference on Web Search and Data Mining}{March 4--8, 2024}{Merida, Mexico}
\acmBooktitle{Proceedings of the 17th ACM International Conference on Web Search and Data Mining (WSDM '24), March 4--8, 2024, Merida, Mexico}
\acmPrice{15.00}
\acmDOI{10.1145/3616855.3635781}
\acmISBN{979-8-4007-0371-3/24/03}

\usepackage{booktabs} 
\usepackage{bm}
\usepackage{multirow}
\usepackage{xspace}
\usepackage[tight,footnotesize]{subfigure}
\usepackage{algorithm} 
\usepackage{amsmath,amsfonts}
\usepackage{enumitem}
\usepackage{algorithmic}
\usepackage{balance}
\usepackage{soul}

\newcommand{\paratitle}[1]{\vspace{1.5ex}\noindent\textbf{#1}}
\newcommand{\ie}{\emph{i.e.,}\xspace}

\newcommand{\eg}{\emph{e.g.,}\xspace}

\newcommand{\wrt}{\emph{w.r.t.}\xspace}
\newcommand{\model}{PEACE}

\begin{document}

\title{PEACE: Prototype lEarning Augmented transferable framework for Cross-domain rEcommendation}

\author{Chunjing Gan}
\authornotemark[1]
\affiliation{
\institution{Ant Group}
\country{}
}
\email{cuibing.gcj@antgroup.com}

\author{Bo Huang}
\authornotemark[1]
\affiliation{
\institution{Ant Group}
\country{}
}
\email{yunpo.hb@antgroup.com}
\thanks{* Equal contribution.}

\author{Binbin Hu}
\affiliation{
\institution{Ant Group}
\country{}
}
\email{bin.hbb@antfin.com}

\author{Jian Ma}
\affiliation{
\institution{Ant Group}
\country{}
}
\email{mj.mj@antgroup.com}

\author{Ziqi Liu}
\affiliation{
\institution{Ant Group}
\country{}
}
\email{ziqiliu@antgroup.com}

\author{Zhiqiang Zhang}
\affiliation{
\institution{Ant Group}
\country{}
}
\email{lingyao.zzq@antfin.com}

\author{Jun Zhou}
\affiliation{
\institution{Ant Group}
\country{}
}
\email{jun.zhoujun@antfin.com}

\author{Guannan Zhang}
\affiliation{
\institution{Ant Group}
\country{}
}
\email{zgn138592@antgroup.com}

\author{Wenliang Zhong}
\authornotemark[2]
\affiliation{
\institution{Ant Group}
\country{}
}
\email{yice.zwl@antgroup.com}
\thanks{$^{\dagger}$ Corresponding author.}

\renewcommand{\shortauthors}{Chunjing Gan et al.}


\begin{abstract}
To help merchants/customers to provide/access a variety of services through miniapps, 
online service platforms have occupied a critical position in the effective content delivery, in which \emph{how to recommend items in the new domain launched by the service provider for customers} has become more urgent.
However, the non-negligible gap between the source and diversified target domains poses a considerable challenge to cross-domain recommendation systems, which often leads to performance bottlenecks in industrial settings.  While entity graphs have the potential to serve as a bridge between domains, rudimentary utilization still fail to distill useful knowledge and
even induce the negative transfer issue.
To this end, we propose \textbf{\model}, a \underline{P}rototype l\underline{E}arning \underline{A}ugmented transferable framework for \underline{C}ross-domain r\underline{E}commendation.
For domain gap bridging, 
{\model} is built upon a multi-interest and entity-oriented pre-training architecture which could not only benefit the learning of generalized knowledge in a multi-granularity manner, but also help leverage more structural information in the entity graph.
Then, we bring the prototype learning into the pre-training over source domains, so that representations of users and items are greatly improved by the contrastive prototype learning module and the prototype enhanced attention mechanism for adaptive knowledge utilization. 
To ease the pressure of online serving, {\model} is carefully deployed in a lightweight manner, and significant performance improvements are observed in both online and offline environments.
\end{abstract}



\maketitle

\input{sec-intro}

\input{sec-rel}

\input{sec-model}
\input{sec-exp}
\input{sec-con}

\balance
\bibliographystyle{ACM-Reference-Format}
\bibliography{references}

\clearpage
\input{sec-app}

\end{document}

%% file: sec-intro.tex
\section{Introduction}
Recently, online service platforms (\eg Alipay, WeChat and Tiktok) have become increasingly prevalent, which attract merchants/customers to provide/access a variety of services through miniapps. 
However, making proper recommendations~\cite{choi2023blurring,sasrec2018,PiBZZG19,neural2017he,lightgcn2020} to satisfy the effective content delivery requirements on real-world service platforms still encounter several intractable issues.
In the case of Alipay, which is known as an integrated platform involving multifarious miniapps, various online providers are able to effortlessly launch new scenarios in the form of miniapps (\eg Travel and Rental). 
Towards new domains, it is 
time-consuming and even impractical to train a recommender from scratch with extremely scarce interactions. 
Fortunately, service platforms generally accumulate 
considerable
global behaviors (\eg search and click on the homepage feed), which shed some light on extracting universal and transferable knowledge for warming up brand-new scenarios.

As a promising direction, cross-domain recommendation (CDR) has attracted a surge of investigations, which enables the effective learning of a data-sparser domain by  transferring useful knowledge 
from data-richer domains.
Existing CDR methods often assume the existence of shared information so that a mapping function can be learned across different domains ~\cite{emcdr2017,commonfeatures2020,dtcdr2019,lu2023three} or availability of source and target domains for joint optimization ~\cite{MMT-Net2020,zhao2022multi,li2023one}. However, this assumption may not hold in real-world applications due to the considerable gap between source and target domains or even unavailability of target domains data during training, which severely hinders the application of existing CDR approaches. 
Recently, several approaches seek to bridge the domain gap via side information such as entity (knowledge) graph and bring the superiority of pre-training~\cite{bert2019,peterrec2020,mae2022} into cross-domain recommendation. Specifically, ~\cite{liu2023pre,gao2023leveraging,tiger2022,wang2021pre,hou2022towards} learn a universal user representation by bridging user's behaviors in multiple source domains via entity graph, text and so on, with the expectation that the obtained universal user representations contain rich information, and can be transferred to various downstream target scenarios directly or with careful fine-tuning. Despite considerable performance improvement in social and e-commerce platforms, rudimentary 
utilization of multiplex knowledge 
may fail to distill useful knowledge and even induce the negative transfer issue in online service platforms integrated with a wide variety of services/miniapps.

Given the above limitations in current approaches, 
we comprehensively explore and effectively exploit rich interactions associated side information in multiple source domains 
to learn universal and transferable knowledge for a variety of unseen downstream domains.
However, the solution is quite non-trivial, which needs to tackle the following essential challenges in industrial environments: 
i) The online service platform includes multifarious scenarios from various service providers with unsharing information and related items without identical properties. Therefore, it is essential to close this gap and utilize multiplex data sources to develope multi-granularity representations to enhance the capability of pre-training.
ii) The prevalence of entity (knowledge) graph in real-world service platforms  encourages us to incorporate it as the bridge for domain gap. However, entity graph in real-world applications is huge, containing 
considerable
ambiguous entities or related entities without identical properties, simply aggregating entity-wise context for universal and transferable knowledge may easily result in a performance bottleneck and render the approach vulnerable to underlying noises.
iii) The scenario-based context that implies varying semantics \wrt different interaction scenarios, plays a vital role in capturing users' underlying preferences, which is expected to be incorporated into user representation in an adaptive manner.

To this end, we propose \textbf{\model}, a \underline{P}rototype l\underline{E}arning \underline{A}ugmented transferable framework for \underline{C}ross-domain r\underline{E}commendation.
In particular, the pre-training stage of {\model} is built upon multi-interest based and entity-oriented architecture, in which users and entities are paired for learning generalized knowledge and multi-granularity representations 
distilled from multiplex interactions in source domains.
To further enhance the capability of learning universal and transferable knowledge, {\model} introduces the prototype learning, so that 
i) the derived contrastive prototype learning component greatly improves the universal knowledge learning by attracting entities with similar characteristics and  repelling unrelated entities;
ii) the user-side representation learning is aware of scenario-based context in an adaptive manner through the proposed prototype enhanced attention mechanism.
In the fine-tuning stage, after effortlessly loading pre-trained parameters, {\model} represents users with the universal encoding and demystifies target items in target domains with specific entities, followed by carefully fine-tuning for 
normal
and 
strictly zero-shot
recommendation in a lightweight manner.

To sum up, we make the following main contributions:
\textbf{i)} We highlight the practical challenges of transferable recommendation in real-world service platforms, which are mainly caused by the large gap between source and target domains, coupled with multiplex knowledge involved. To our knowledge, we take the first step to employ prototype learning for cross-domain recommendation for the extraction of universal and transferable knowledge.
\textbf{ii)} We devise a novel transferable framework {\model} in a multi-interest based and entity-oriented pre-training manner, which is equipped with contrastive prototype learning and prototype enhanced attention to substantially facilitate the universal knowledge learning for entities and the scenario-aware representation for users, respectively. Moreover, to ease the pressure of online serving, {\model} is carefully deployed in a lightweight manner.
\textbf{iii)} We conduct extensive experiments in both offline and online environments, which demonstrate that {\model} significantly surpasses a series of state-of-the-art approaches for cross-domain recommendation in both normal and zero-shot setting. And in-depth analysis also reveals the superiority of prototype learning based components, with which the distance of similar entities in the representation space can be potentially narrowed down.

%% file: sec-rel.tex
\section{Related work}

\paratitle{Graph based Recommendation}
Recently, a surge of 
attention has been dedicated to graph based recommendation, which flexibly characterizes various kinds of auxiliary context (\eg collaborative graphs~\cite{wang2019neural,lightgcn2020}, social graphs~\cite{fan2019graph,yu2021self}, knowledge graphs~\cite{wang2019kgat,huang2021knowledge} and heterogeneous graphs~\cite{hu2018leveraging,lu2020meta}) for facilitating recommendation.
Early studies carry structural information based on extracting and encoding of numerous paths in graphs~\cite{wang2019explainable,hu2018leveraging}. 
By taking full advantage of the information propagation, graph neural networks~\cite{huai2023m2gnn,wei2023lightgt,wu2020comprehensive} help graph-based recommenders explicitly model the high-order connectivities in graphs in an end-to-end fashion~\cite{wu2022graph}.
 More recently, several works encourage the independence of different user intents/interests behind graphs in representation learning for better model capability~\cite{wang2021learning,disengnn2022}.
Although these methods usually achieve promising performance,
they either require the availability of data from both source and target domains or cannot bridge isolated items from different domains, especially for unknown items in anonymous target domains, thus failing to make transferable recommendation in real-world service platforms when target domain items are unavailable.


\paratitle{Cross-domain Recommendation}
Cross-domain recommendation~\cite{cdrsurvey2021,cdrsurvey2022} aims at alleviating the long-standing cold-start issue in recommendation by leveraging the rich interactions in source domains to help the learning of recommenders in target domains.
A series of studies propose to learn a shared mapping matrix, and then perform alignment and transformation between source and target users/items~\cite{emcdr2017,semitrans2019,dtcdr2019,ddtcdr2020,bitrans2020}.
Recently, due to the impressive and rapid adaptability of meta learning to new tasks with only few data, recent attentions have been shifted towards the effective incorporation of meta networks, which focus on bridging user preference in different domains~\cite{persontrans2022}, including TMCDR~\cite{tmcdr2021} that is based on  the popular Model-Agnostic Meta-Learning (MAML) and PTUPCDR~\cite{persontrans2022} that models personalized bridge with users' preferences and characteristics. 
However, all the methods above are incapable of leveraging graph structural information to enhance the representation of both source and target domains.
Moreover, most of the current approaches commonly involve interactions of target domains in the training process, which ignore a practical fact in real-world service platforms that items are completely non-overlapping across different domains and items in target domains are totally unavailable in the training stage.

\paratitle{Pre-training for Recommendation}
Given the unprecedented success in learning useful representations with self-supervised signals, the pre-training technique has attracted a surge of investigation on recommendations. 
Relying on the recent development of self-supervised learning, many efforts have been made to pre-train a powerful recommender based on various auxiliary tasks, including sequence-level masking~\cite{ma2020disentangled,bert4rec2019,gu2021self}, item-level~\cite{zhou2021contrastive,yao2021self} and graph-level~\cite{wu2021self,simgcl2022} contrastive learning.
In terms of the pre-training for cross-domain recommendation, several works have attempted to learn transferable interest based on sequence~\cite{peterrec2020}, collaborative graph~\cite{wang2021pre} and  knowledge graph~\cite{tiger2022}. 
However, existing methods cannot handle the large gap between source and target domains, coupled with multiplex behaviors across domains. Although knowledge graphs are introduced in  current approaches to connect isolated items, the entity-wise context is still insufficient to learn universal and transferable knowledge with the huge and noisy graph in real-world service platforms.

%% file: sec-model.tex
\begin{figure}[!t]
    \centering
    \includegraphics[width=1.0\columnwidth]{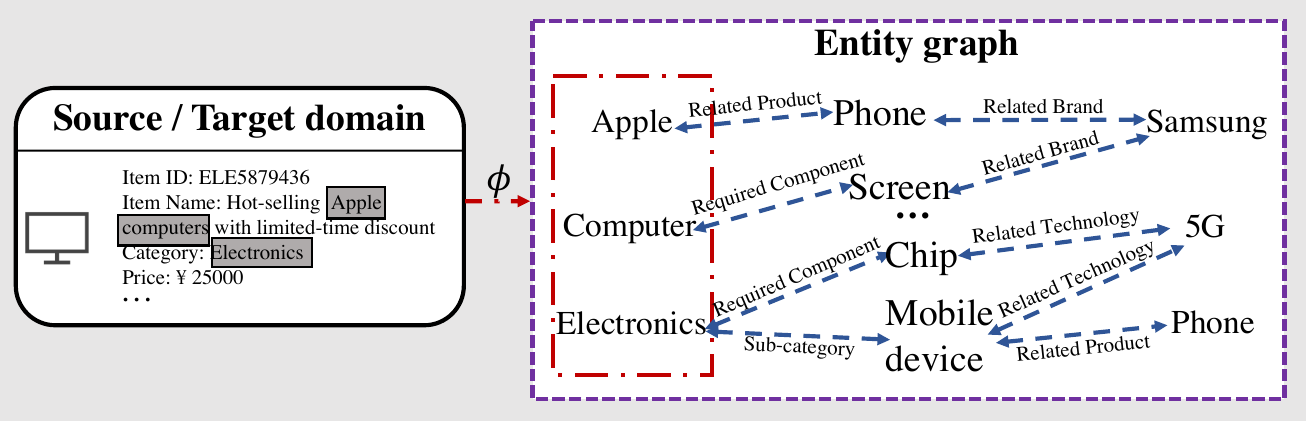}
    \caption{The overall procedure of item-entity mapping and graph reasoning.}
    \label{fig:mapping2}
\end{figure}

\section{Preliminaries} 
Let $\mathbb{S}$ and $\mathbb{T}$ denote the set of source and target domains, 
we define each source domain as $\mathcal{S} \in \mathbb{S}$ with the user set $\mathcal{U}^\mathcal{S}$ and the item set $\mathcal{I}^\mathcal{S}$, comprised of interaction records $\mathcal{H}^\mathcal{S} = \{u, i, \mathcal{B}_{ui}, y_{ui} | u \in \mathcal{U}^\mathcal{S}, i \in \mathcal{I}^\mathcal{S}\}$. Here, $\mathcal{B}_{ui} \subset \mathcal{I}^\mathcal{S}$ denotes historical behaviors (\ie item list) for user $u$ when item $i$ is recommended and $y_{ui} \in \{0, 1\}$ is the feedback of user \wrt item i (\eg $y_{ui} = 1$ when clicking, 
otherwise $0$).
Analogically, we define each target domain as $\mathcal{T} \in \mathbb{T}$ with the user set $\mathcal{U}^\mathcal{T}$ and the item set $\mathcal{I}^\mathcal{T}$, containing interactions records $\mathcal{H}^\mathcal{T} = \{u, i, \mathcal{B}_{ui}, y_{ui} | u \in \mathcal{U}^\mathcal{T}, i \in \mathcal{I}^\mathcal{T}\}$, where  $\mathcal{B}_{ui} \subset \mathcal{I}^\mathcal{T}$.  

The goal of our task is to comprehensively explore and flexibly exploit universal knowledge from multiple source domains $\mathbb{S}$, and then perform knowledge transfer for facilitating recommendation in each target domain $\mathcal{T} \in \mathbb{T}$. Since new scenarios are commonly launched in  real-world service platforms, the  transferable recommendation task may have following essential properties:
i) Interactions in target domains are extremely scarce (\ie $|\mathcal{H}^\mathcal{T}| \ll |\mathcal{H}^\mathcal{S}|$) and even totally unavailable (\ie $\mathcal{H}^\mathcal{T} = \emptyset$), which presses for enough universal and transferable knowledge without noises. 
ii) Completely non-overlapping items across domains (\ie $\mathcal{I}^{\mathcal{S}} \cap \mathcal{I}^{\mathcal{T}} =  \emptyset$) encourage our approach to perform item-level alignment with inductive ability.
iii) Anonymity of target domains means the procedure of model training cannot involve any target domain $\mathcal{T} \in \mathbb{T}$, where transferable capability with fast adaptation is urgent.

To effectively bridge the domain gap in multiple scenarios, we naturally leverage the generality of entity graph \cite{kddyiru}, denoted as $\mathcal{G} = \left \{ \mathcal{E}, \mathcal{R} \right \}$
with an entity set $\mathcal{E}$ and a relation set $\mathcal{R}$, to encode universal user/item representation. Specifically, \cite{kddyiru} is extracted from contents/items on the platform of Alipay\footnote{https://www.alipay.com} and its construction pipeline is a mixture of multiple text mining techniques (\eg NER~\cite{lample2016ner} and RE~\cite{Zhu2019re}).
To embrace broader applications in multiple tasks such as recommender system in online service platforms where items are the basic existence forms, \cite{kddyiru} provides a unified API to associate entities with items, \ie $\phi : \mathcal{I} \rightarrow \mathcal{E}$~\footnote{For brevity, we
denote $\mathcal{I}_{\mathcal{S}}$ or  $\mathcal{I}_{\mathcal{T}}$ as $\mathcal{I}$. Moreover, $\phi(\cdot)$ 
can
map one item to multiple entities and in our experiments, for each item we set the maximum associated entity number to 3 \wrt the average entity number per item in our platform, which is 2.78.}. 
A toy example of mapping between items and entities and the following entity graph reasoning process is illustrated in Figure \ref{fig:mapping2}.

\section{The proposed approach}
\begin{figure}[!t]
    \centering
    \includegraphics[width=1\columnwidth]{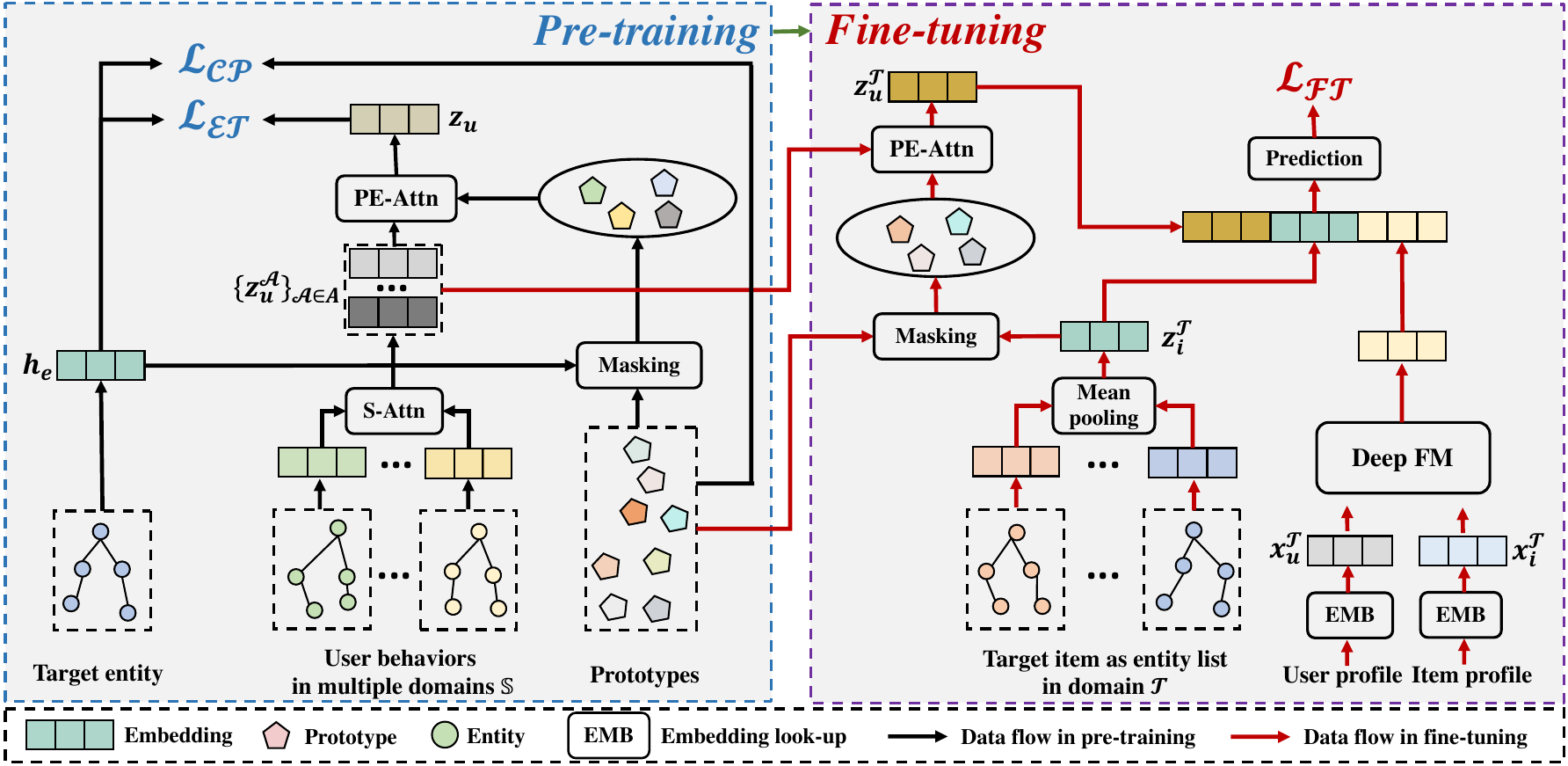}
    \caption{Overall architecture of {\model}, where S-Attn and PE-Attn denote self-attention and prototype enhanced attention.}
    \label{fig:model}
\end{figure}

In this section, we present {\model}, a domain-agnostic cross-domain framework, whose overall architecture is presented in Figure ~\ref{fig:model}.

\subsection{Multi-interest based and Entity-oriented Recommender Pre-training} \label{sec:entity_pretrain}
Essentially, the online service platform is a composite of multifarious scenarios launched by a variety of service providers, where information is commonly unshared, making related items (\eg same categories and brands) without identical properties. 
To help close the distances and effectively leverage multiplex interactions, we draw support from the well-established entity graph to characterize multiple interests with entity-wise context for generalized recommender pre-training.
Firstly, taking full advantage of information propagation, we perform graph convolution operation~\cite{gat2018} over the entity graph to encode entities in the entity graph $\mathcal{G}$ as the $d$-dimensional matrix $\mathbf{H}^{\mathcal{G}} =\{\bm{h}_e\}_{e \in \mathcal{E}} \in \mathbb{R}^{|\mathcal{E}| \times d}$.

\subsubsection{Hierarchical-attentive User Representation across Domains}
To allow the entity graph to serve as the critical ``bridge'' across multiple domains, we propose to deconstruct the user behaviors in each domain into a series of entity list, which could be utilized to comprehensively characterize user interests across domains in a unified manner.
Given a user $u$ in source domain $\mathcal{S}$, the deconstruction process with the item-entity alignment function $\phi$  go on as follows:
$User \xrightarrow{\mathcal{B}_{u,i}} Item\ list \xrightarrow{\phi} Entity\ list$.
With straightforward retrieval from the entity representations $\mathbf{H}^{\mathcal{G}}$, we can obtain the representation for user $u$ in domain $\mathcal{S}$, denoted as the matrix $\mathbf{H}_u^{\mathcal{S}} \in \mathbb{R}^{H \times d}$, where $H$ is the maximum sequence length in each domains. To comprehensively understand user's interests among multiple source domains, 
we use a hierarchical-attentive mechanism to effectively absorb universal and transferable knowledge into user encodings.

\paratitle{Entity encodings $\rightarrow$ Interest-specific user encodings.}
Given the representation of user $u$ in multiple domains $\mathbb{S}$, expressed as a sequence set of entity representations derived from graph learning (\ie $\{\mathbf{H}_u^{\mathcal{S}}\}_{\mathcal{S} \in \mathbb{S}}$), we naturally perform multi-interest exploration with a sequence encoder. 
Inspired by the compelling design and superior performance of Transformer~\cite{transformer2017}, 
we adopt a self-attention layer with $M$ interest kernels (\ie $\{\bm{w}_1, \cdots, \bm{w}_M\}$) for multi-interest extraction.  
Specifically, the sequence encoder starts with concating entity representations of user $u$ from multiple domains as $\mathbf{H}_u = ||_{\mathcal{S} \in S}\mathbf{H}_u^{\mathcal{S}} \in \mathbb{R}^{H|\mathbb{S}| \times d}$, and extracting the $i$-th interest as follows:
\begin{equation} 
     \label{eq:domain-specific}
    \bm{z}^{(i)}_u = \text{SoftMax}(\bm{w}_i^T tanh(\mathbf{W}^{\mathcal{A}} \cdot \mathbf{H}_u^T)) \cdot \mathbf{H}_u^T,
\end{equation}
where $\mathbf{W}^{\mathcal{A}}$ is the weight matrix for interest extraction. After $M$ times of interest extraction, we re-denote the all obtained interest representations as $\{\bm{z}_u^{\mathcal{A}}\}_{\mathcal{A} \in \mathbb{A}}$, where $\mathbb{A}$ is the interest set.

\paratitle{Interest-specific encodings $\rightarrow$ Universal encodings.}
Here, we dive into the integration of user encodings derived from multiple interests from various source domains, and at the same time characterizing the fine-grained differentiation of importance or relevance across domains. 
Given the interest-specific representation set $\{\bm{z}^{\mathcal{A}}_u\}_{\mathcal{A} \in \mathbb{A}}$, we obtain the universal encoding for user $u$ with adaptive fusion as follows:
\begin{equation} 
\label{eq:universal}
    \bm{z}_u = \sum_{\mathcal{A} \in \mathbb{A}} \text{SoftMax}({\bm{v}^T(\tanh(\mathbf{W} \cdot \bm{z}^{\mathcal{A}}_u))}) \cdot  \bm{z}^{\mathcal{A}}_u, 
\end{equation}
where $\bm{v}$ and $\mathbf{W}$ are the weight vector and matrix, respectively.

\subsubsection{Entity-oriented Recommender Pre-training}
As mentioned above, items between source and target domains are unshared, making it 
infeasible to pre-train a recommender aiming at the matching between user interests and specific items.
In addition, entities usually involve more generalized knowledge (\ie brand and category), which potentially benefits the transferable recommendation. 
Hence, in this paper, we adopt an entity-oriented pre-training strategy to explore and exploit universal knowledge in multiple source domains.
Firstly, given a record $\langle u, i, y_{ui}\rangle  \in \mathcal{H}^S$  in domain $\mathcal{S}$~\footnote{For simplicity, we omit $\mathcal{B}_{ui}$ in the following.}, with the help of the alignment function $\phi(\cdot)$, we map it to the entity-level forms $\langle u, e, y_{ue}\rangle $ with $e = \phi(i)$ and $ y_{ue} =  y_{ui}$. It is noteworthy that one original record would be mapped to multiple entity-level forms since one item may be associated with multiple entities. 
In sum, we organize the  entity-level form as $\hat{\mathcal{H}}^{\mathcal{S}} = \{u, e, y_{ue} | u \in \mathcal{U}^{\mathcal{S}}, e \in \mathcal{E}\}$. 
For each training pair $\langle u, e \rangle $, we could easily obtain the corresponding representations $\bm{z}_{u}$ and $\bm{h}_e$, and adopt a multi-layer perceptron based decoder (\ie MLP($\cdot$)) for prediction, followed by cross-entropy (\ie $\mathcal{C}(\cdot, \cdot)$) based optimization.
\begin{equation}
        \hat{y}_{ue} = \sigma(\text{MLP}(\bm{z}_{u}||\bm{h}_e)),
\end{equation}
\begin{equation}
    \label{eq:loss_et}
        \mathcal{L}_{\mathcal{ET}} = \sum_{\mathcal{S} \in \mathbb{S}}\sum_{\langle u, i, y_{ue}\rangle  \in \hat{\mathcal{H}}^{\mathcal{S}}} \mathcal{C}(y_{ue}, \hat{y}_{ue}),
\end{equation}
where ``$||$'' is the concatenation operation and $\sigma(\cdot)$ is the sigmoid function. In industrial scenarios, the huge scale of graphs and the  latency requirements usually restrict the recommender system to 2-hop graph learning, which makes the entity-oriented pre-training strategy more effective to learn universal knowledge from entity graph, \ie entity-oriented strategy includes more structural neighbors (``entity $\rightarrow$ entity $\rightarrow$ entity'' $v.s.$ ``item $\rightarrow$ entity $\rightarrow$ entity'').

\subsection{Improving Entity Representations with Contrastive Prototype Learning} \label{sec:prototype}
Until now, we have introduced the entity-oriented recommender pre-training which characterizes entity-wise context for learning universal and transferable knowledge that potentially benefits various downstream applications. Although the entity-wise context provides excellent capability for generalizing in the transferable recommendation, we can further improve the universal knowledge learning by attracting  entities with similar characteristics (\eg ``Luckin'' and ``Starbucks'') and repelling unrelated entities (\eg ``Starbucks'' and ``Hot pot''). Inspired by 
cluster analysis\cite{kmeans++,aggarwal2013}, we wish to introduce the concept of prototype into the recommender pre-training for helping cluster semantically similar entities in the representation space, so that 
the pre-trained recommender would be not only more generalized with cluster-level semantic, but also invulnerable to inevitable noises in real-world environments.

To this end, we propose a contrastive prototype learning module to improve entity representation. 
Specifically, we adopt a $d$-dimensional learnable matrix $\mathbf{P} \in \mathbb{R}^{n \times d}$ to denote the latent prototypes in the entity graph, where $n$ is the number of prototypes. 
For each entity $e$, we calculate the normalized similarity vector \wrt the $n$ prototypes as follows:
\begin{equation}
    \bm{s}_e = \text{SoftMax}(\mathbf{P} \cdot \bm{h}_e).
    \label{eq:sim}
\end{equation}

To well guide the learning process of prototypes,  we wish the related prototype could help ``guess'' the target entity based on the weighted sum over the similarity distribution. Hence, given the target entity $e$, we obtain its prototype-level view as follows:
\begin{equation}
\label{eq:pt_view}
    \hat{\bm{h}}_e = \mathbf{P}^T \cdot \bm{s}_e.
\end{equation}

After that, the goal of the proposed contrastive prototype learning is to pull related views close and push others away, we define the loss based on InfoNCE~\cite{infonce2018} as follows:
\begin{equation}
 \label{eq:loss_cp}
    \begin{split}
    \mathcal{L}_{\mathcal{CP}} &= -\sum_{e \in \mathcal{E}} \left(  \log{\frac{\exp{(\delta(\bm{h}_e, \hat{\bm{h}}_e)/\tau)}}{\sum_{e' \sim \mathcal{P}_{neg}}\exp{(\delta(\bm{h}_e, \hat{\bm{h}}_{e'})/\tau)}}} \right. \\
    & \left. + \log{\frac{\exp{(\delta(\hat{\bm{h}}_e, \bm{h}_e)/\tau)}}{\sum_{e' \sim \mathcal{P}_{neg}}\exp{(\delta( \hat{\bm{h}}_e, \bm{h}_{e'})/\tau)}}}  \right), \\
    \end{split}
\end{equation}
where $\mathcal{P}_{neg}$ is the noise distribution for generating negative views, which could be set as uniform distribution over the current batch for efficient training or other biased distributions for hard negatives. And $\tau$ is the temperature parameter that controls the strength of penalties we enforced on hard negative samples~\cite{wang2021under} and $\delta(\cdot, \cdot)$ measures the similarity of two target views with cosine function.

\subsection{Improving User Representations with Prototype Enhanced Attention} \label{sec:prototype_enhanced_attention}
In the pre-training stage, data collections in source domains involve rich user behaviors in various scenarios,  \ie visiting a travel scene with travelling plan while visiting a job hunting scene when seeking employment. However, aforementioned process of universal user encoding (Eq.~\ref{eq:domain-specific} and Eq.~\ref{eq:universal}) does not take the user and scenario-based context into consideration, which lacks the ability to capture varying semantics \wrt different interaction scenarios. 
Hence, we aim at improving user representations by facilitating the hierarchical-attentive mechanism with prototype enhanced attention.

Formally, given the target user-entity pair $\langle u, e\rangle$ in the source domain $\mathcal{S}$, we could easily obtain the similarity distribution of target $e$ towards $n$ prototypes as $\bm{s}$ through Eq.~\ref{eq:sim}. Then we use a mask vector $\bm{m} \in \mathbb{R}^{n}$ to keep the most relevant prototypes \wrt target entity $e$, whose $i$-th value can be defined as follows:
\begin{flalign} \label{eq:mask}
    \bm{m}_i &=\left\{\begin{array}{ll}
    0, & \bm{s}_i  \ge \epsilon(\bm{s}, K); \\
    -\infty, & \text { otherwise, }
    \end{array}\right.
\end{flalign}\noindent
where $ \epsilon(\bm{s}, K)$ aims at finding the $K$th-largest entries from $\bm{s}$. 

Since prototypes serve as informative centers in the entity graph, we demystify the scenario-based context with most relevant entities and capture varying semantics \wrt different related prototypes. 
In particular, given the interest-specific representation of user $u$  as $\bm{z}_u^{\mathcal{A}}$ (See Eq.~\ref{eq:domain-specific}), we produce the scenario-aware representation by weighing various underlying preferences among the target user $u$ and prototypes $\mathbf{P}$ as follows:
\begin{equation}  
    \label{eq:hatz}
        \hat{\bm{z}}_u^{\mathcal{A}} =  \text{MLP}(\mathbf{P}^T \cdot \text{SoftMax}(\frac{\mathbf{P} \cdot \bm{z}_u^{\mathcal{A}} + \bm{m}_i}{\sqrt{d}}) +  \bm{z}_u^{\mathcal{A}}). \\
\end{equation}
With the improved interest-specific representation, we feed it into Eq.~\ref{eq:universal} for universal encoding by replacing the original interest-specific representation set $\{\bm{z}_u^{\mathcal{A}}\}_{\mathcal{A} \in \mathbb{A}}$ with 
$\{\hat{\bm{z}}_u^{\mathcal{A}}\}_{\mathcal{A} \in \mathbb{A}}$. After that, we re-denote the improved universal encoding for user $u$ as $\hat{\bm{z}}_u$.

\begin{figure}[t]
    \centering
    \includegraphics[width=0.95\columnwidth]{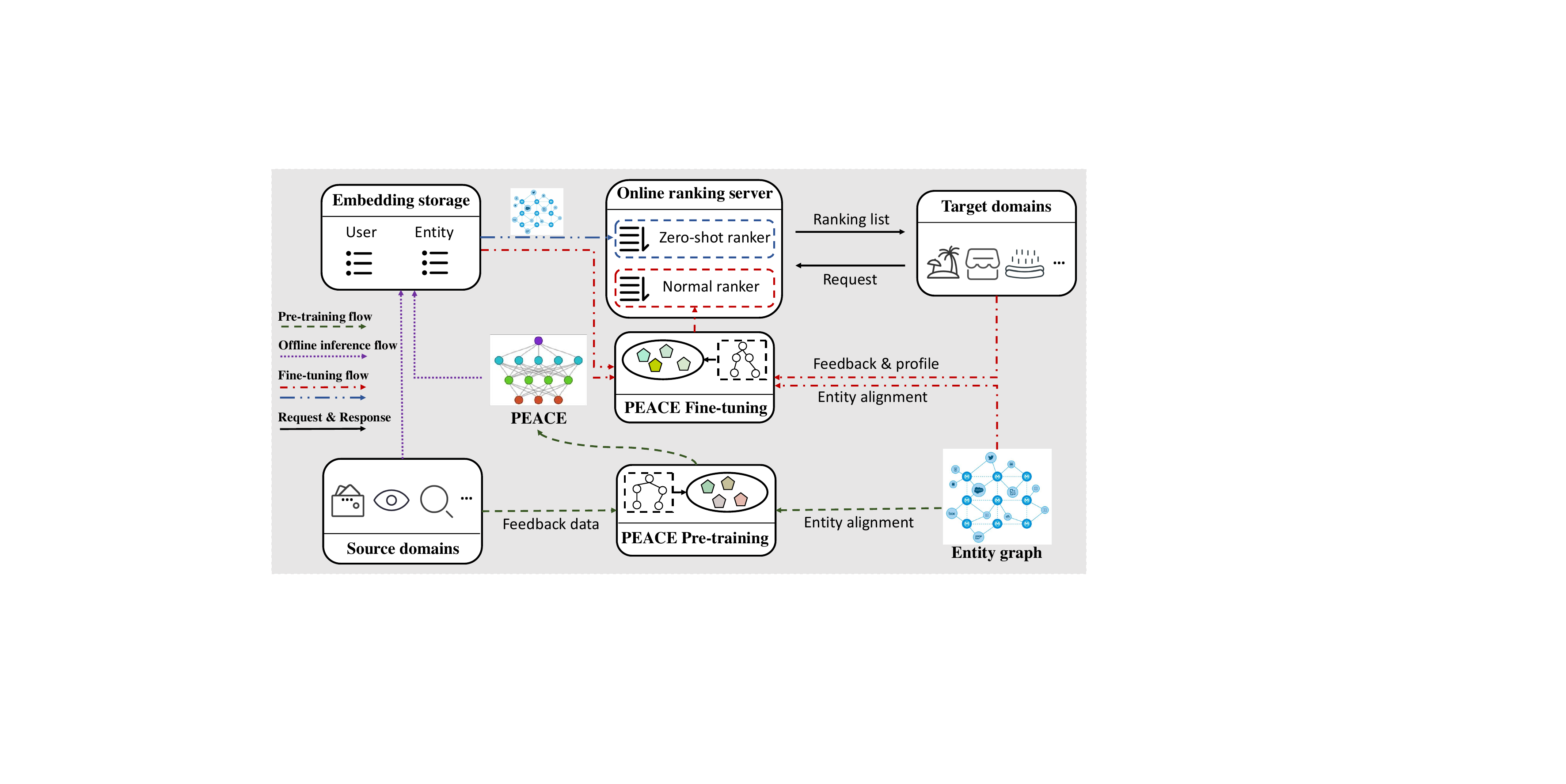}
    \caption{The system architecture of {\model}.}
    \label{fig:system}
\end{figure}

\subsection{Learning of {\model}} \label{sec:learning}
\paratitle{Pre-training Stage in Source Domains}
By integrating entity-oriented pre-training strategy with contrastive prototype learning module, the overall objective can be defined as follows:
\begin{equation}
    \label{eq:loss_source}
    \mathcal{L}_{\mathcal{PT}} = \mathcal{L}_{\mathcal{ET}} + \gamma \cdot \mathcal{L}_{\mathcal{CP}}
\end{equation}
where $\gamma \geq 0$ denotes the relative
weight of contrastive term.

\paratitle{Fine-tuning Stage in Target Domains}
The entity-oriented pre-training strategy endows our approach superior generalization ability for making fast adaptation to various target domains, since there is no need to care about anonymous items in target domains.    

Formally, with a target domain $\mathcal{T} \in \mathbb{T}$, we easily initialize a recommender by utilizing the pre-training parameters, which is capable of producing representations of users and entities. 
And then, given a user-item pair $\langle u, i\rangle$ in target domain $\mathcal{T}$, the user representation could be naturally obtained through universal encoding (denoted as $\bm{z}^\mathcal{T}_u$), while we obtain the final representation for item $i$ (denoted as $\bm{z}^\mathcal{T}_i$) by adopting a mean pooling operation~\footnote{The straightforward mean pooling operation helps avoid amounts of delays for online inference, which could be naturally extended to complicated attentive aggregation.} on the representations of entities associated with item $i$.
Thus, the preference score is given by MLP and DeepFM~\cite{deepfm2017} as follows:
\begin{equation} \label{eq:y_target}
    \hat{y}_{ui} = \sigma(\text{MLP}(\bm{z}^\mathcal{T}_u || \bm{z}^\mathcal{T}_i||\text{DeepFM}(\bm{x}^\mathcal{T}_u||\bm{x}^\mathcal{T}_i))),
\end{equation}
where $\bm{x}^\mathcal{T}_u$ and $\bm{x}^\mathcal{T}_i$ denote the profile information of user $u$ and item $i$ in target domain $\mathcal{T}$, respectively.  
Similarly, we adopt a  cross entropy loss as the final objective:
\begin{equation}  \label{eq:loss_target}
    \mathcal{L}_{\mathcal{FT}} = \sum_{\langle u, i, y_{ui}\rangle \in \mathcal{H}^{\mathcal{T}}} \mathcal{C}(y_{ui}, \hat{y}_{ui}).
\end{equation}
In general, the fine-tuning stage can be 
extended to consider the CVR estimation~\cite{esm22020} by adopting the ESSM-like architecture~\cite{esmm}. 
Our approach also can 
effortlessly adapt to strictly cold-start scenarios (\ie with no supervised/labeled data for some target domains), which are common in online service platforms, by calculating the preference score with the inner product of $\bm{z}^\mathcal{T}_u$ and $\bm{z}^\mathcal{T}_i$ and encouraging results are presented in the ``Experiments'' part.

\subsection{Lightweight Deployment of {\model}} \label{sec:deployment}
To ease the pressure of online serving, we strive to deploy the proposed {\model} in a lightweight manner. 
We present the online deployment pipeline of {\model} in Figure~\ref{fig:system}, which firstly cold start a new scenario and then fine-tune an expressive recommender with accumulated interactions. 
In particular, the deployment of {\model} mainly contains three central flows:
\begin{itemize}[leftmargin=*]
    \item \emph{Pre-training flow.} 
    With the collected feedback data from multiple domains, coupled with a well-established entity graph, the proposed {\model} is pre-trained daily to keep universal and transferable knowledge stay up-to-date.
    \item \emph{Offline inference flow.} 
    To ease the burden graph neural network imposes on online serving system, we aim at decoupling the entire procedure by inferring embeddings for users and entities in advance. With the knowledge in \emph{Embedding Storage}, only MLP based components are fine-tuned by {\model} for downstream applications, so that the complicated information propagation is unnecessary, which greatly eases the pressure of online serving.
    \item \emph{Fine-tuning flow.} 
    As no interaction is available when a new domain is launched by the service provider, {\model} provides recommendation services with the following two steps: 
    i) A zero-shot ranker that helps to cold start the scenario with the direct inner product via embeddings of users and items.
    ii) With accumulated interactions, as well as the profile data, {\model} would load the pre-trained parameters for fine-tuning, and then upload the well-trained model to the 
    normal ranker.
\end{itemize}

%% file: sec-exp.tex
\section{Experiments}

\input{tab/data_tab.tex}

\input{tab/main_tab.tex}
\subsection{Experimental Setup}
 
\subsubsection{Datasets}
Since the proposed {\model} mainly focuses on the industrial cross-domain recommendation issue in online service platforms, we evaluate it by employing real-world industrial datasets\footnote{The dataset does not contain any Personalized Identifiable Information.} which are collected from different domains in Alipay. 
Specifically, for the source domains, we collect one month's data from \textbf{Payment}, \textbf{Search} and click behaviors on the \textbf{Homepage Feed}. 
In terms of the target domains, due to data sparsity, we collect two month's data from the following six scenarios: \textbf{Rental} service, \textbf{Travel} service, \textbf{Food Delivery} service, \textbf{Digital Art} service, 
\textbf{Dining} service, and \textbf{Daily Necessity} service.
Moreover, an entity
graph involving more than ten millions of entities and more than one hundred millions of triplets is incorporated for cross-domain recommendation, where we extract a 32-dimensional embedding from a pre-trained BERT for each entity as the original features. In each target domain, several attributes are also extracted for model training, \ie user-side attributes include age, occupation, behaviors in current domains and so on while item-side attributes contain item id, category, brand and so on. Detailed statistics of our datasets are shown in
Table~\ref{tab:data_sta}.

\input{tab/main_table_zero_shot.tex}

\subsubsection{Baselines}
We compare {\model} with following SOTAs:

\begin{itemize}[leftmargin=*]
    \item \textbf{Basic competitors}: We train a classical \textbf{DeepFM}~\cite{deepfm2017} for each target domain with basic features of users and items.

    \item \textbf{Pre-trained competitors}: Similar as {\model}, we pre-train a transferable and expressive model on multiple source domains, and then fine-tune\footnote{Basic features of users and items are involved via DeepFM as in Eq. \ref{eq:y_target}.} it on each downstream domain. They include \textbf{Sequential learning}, following the BERT-like pre-training strategy (\ie \textbf{SASRec}~\cite{sasrec2018}, \textbf{BERT4Rec}~\cite{bert4rec2019}, \textbf{PAUP}~\cite{paup2022}, \textbf{PeterRec}~\cite{peterrec2020} and \textbf{ComiRec}~\cite{comirec2020}), \textbf{Contrastive graph learning} ( \ie \textbf{SimGCL}~\cite{simgcl2022}) and \textbf{Cross-domain graph learning} (\ie \textbf{PCRec}~\cite{wang2021pre} with LightGCN \cite{lightgcn2020} and Disen-GNN \cite{disengnn2022} as the backbones, denoted as \textbf{PCRec-L} and \textbf{PCRec-D}, respectively).

    \item \textbf{Competitors for zero-shot recommendation}: We evaluate the performance of {\model} in strictly zero-shot (\ie cold-start) scenarios, where the state-of-the-art \textbf{Tiger}~\cite{tiger2022} is the competitor.
\end{itemize}

We discuss the working details in the supplemental material, and adopt  two widely-used metrics, \ie \textbf{Hit} and \textbf{NDCG} in the experiments (we truncate the ranking list to 5 and 10). 

\subsection{Overall Performance}

From the empirical results on the transferable recommendation task in normal (Table~\ref{tab:overall_result2}) and zero-shot (Table~\ref{tab:zero_shot_result}) settings, the major findings can be summarized as follows:
\begin{itemize}[leftmargin=*]
\item \textbf{Superior capability of {\model} for transferable recommendation.} 
We observe significant performance gains on six target domains in both normal and zero-shot settings,
demonstrating the remarkable strength of {\model} in characterizing universal and transferable knowledge.
The substantial improvements \wrt Tiger in zero-shot recommendation further demonstrate the transferable capability of {\model}, which could greatly warm up a recommender without fine-tuning. 
\item \textbf{Transferable recommendation benefits from pre-training manner.} 
In most cases, pre-training $\&$ fine-tuning models beat the basic competitor, indicating the usefulness of rich behaviors in source domains, and at the same time suggesting that a well-defined pre-trained model can help the recommenders in target domains. However, we also notice that several pre-trained models perform worse in a few domains, \ie so-called negative transfer~\cite{negativetransfer2019}, which are alleviated in {\model} by abstracting and incorporating prototype-level knowledge.
\item \textbf{Prototype learning helps avoid noise in entity graph.}
Generally, we find graph learning approaches based on contrastive signals and the cross-domain framework are both unable to yield promising performance, which is probably attributed to inevitable noises derived from  the huge entity graph in real-world applications. Fortunately, our {\model} proposes to characterize prototype-level knowledge in a fine-grained manner, which renders it invulnerable to unsatisfying noises. 
\item \textbf{Entity-oriented pre-training is better.}
Surprisingly, when compared with graph learning approaches, pre-trained sequential learning based approaches often achieve better performance, notably BERT4Rec and PAUP. Such improvements mainly lie in the compelling design of Transformer architecture, more importantly, the entity-oriented pre-training strategy, which directly matches entity-level semantic over sequence for more universal knowledge towards transferable recommendation.
\end{itemize}

\subsection{Further Probe~\label{sec:in-depth-ana}}
In this section, we make a series of in-depth analysis to better understand the virtues of the proposed {\model}.
In the following parts, we only present the performance comparison results \wrt the NDCG metric, and the similar trends could be observed under the Hit metric. Detailed comparison results are provided in the supplemental material.

\subsubsection{Ablation Study}
\begin{figure}
	\centering
	\subfigure[NDCG@5]{\includegraphics[width=0.45\columnwidth]{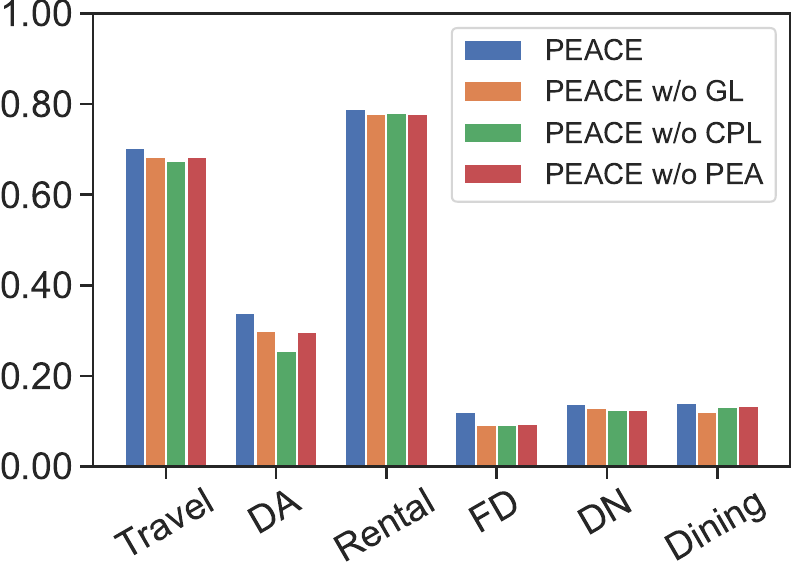}}
	\subfigure[NDCG@10]{\includegraphics[width=0.45\columnwidth]{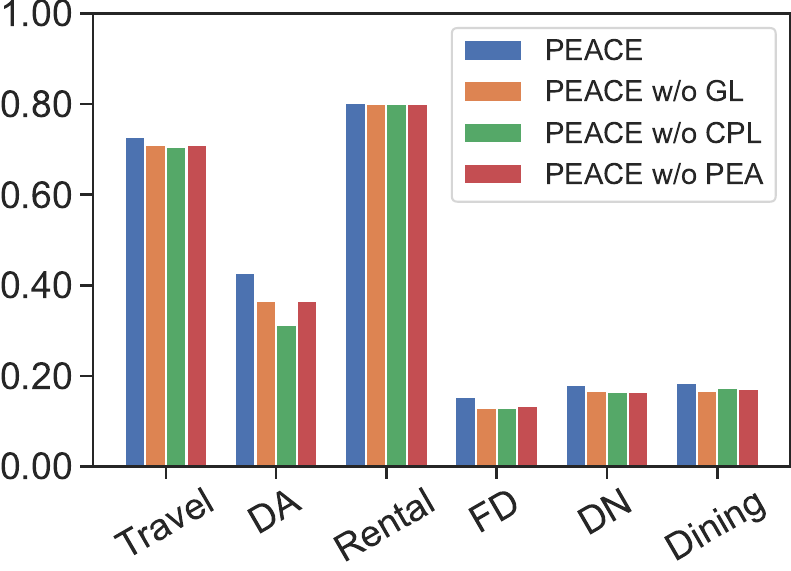}}
  \caption{Ablation study of {\model}. DA, FD and DN denote Digital Art, Food Delivery and Daily Necessity.}
	\label{fig:ab_study}
\end{figure}

We examine the effectiveness of each 
component in {\model} by preparing following variants:
i)  \textbf{{\model} w/o GL} (removing the \textbf{G}raph \textbf{L}earning component for entity encoding),
ii) \textbf{{\model} w/o CPL} (removing \textbf{C}ontrastive \textbf{P}rototype \textbf{L}earning component),
and iii)  \textbf{{\model} w/o PEA} (removing \textbf{P}rototype \textbf{E}nhanced \textbf{A}ttention).
We plot the performance comparison results in Figure~\ref{fig:ab_study}, 
which suggests
the performance would drop a lot when one component is discarded. 
It is worthwhile to note that {\model} w/o CPL performs worst in comparison, which reveals the critical position of prototype learning in capturing universal and transferable knowledge to benefit downstream domains.

\subsubsection{Qualitative Analysis of Contrastive Prototype Learning}
\begin{figure}
	\centering
	\subfigure[{\model} w/o CPL]{\includegraphics[width=0.48\columnwidth]{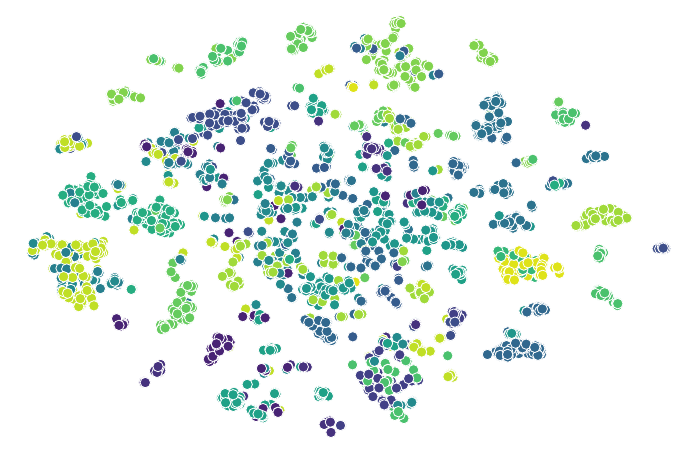}}
        \subfigure[{\model}]{\includegraphics[width=0.48\columnwidth]{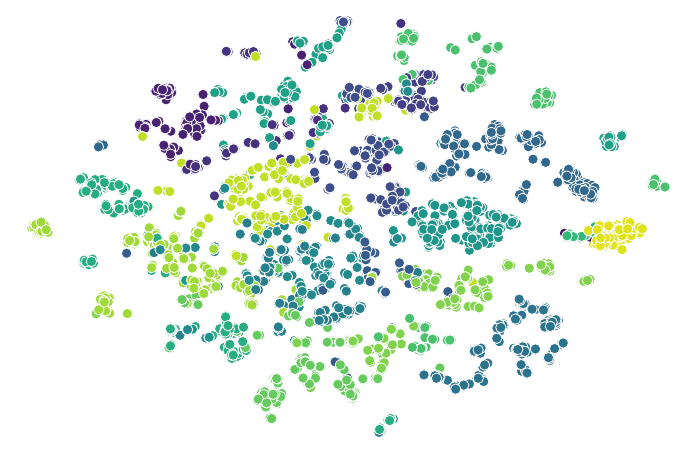}}
  \caption{The 2-D visualization (with t-SNE~\cite{tsne2008}) of entities,  where the color represent the corresponding prototype.}
	\label{fig:visualization_300}
\end{figure}

Firstly, we randomly select 6000 entities in the entity graph, and visualize the learned entity embedding from {\model} w/o CPL and {\model}  using t-SNE in Figure~\ref{fig:visualization_300}, where different colors denote the prototypes that each entity belongs to.
We observe when compared with {\model} w/o CPL, the entity clusters from {\model} is more coherent \wrt prototypes. It suggests that the contrastive prototype learning and the derived prototypes, could help close the distance of similar entities in the representation space, and 
improve
performance of {\model} by better capturing noise-free and universal knowledge.

\begin{figure}
	\centering
	\subfigure[Travel]{\includegraphics[width=0.32\columnwidth]{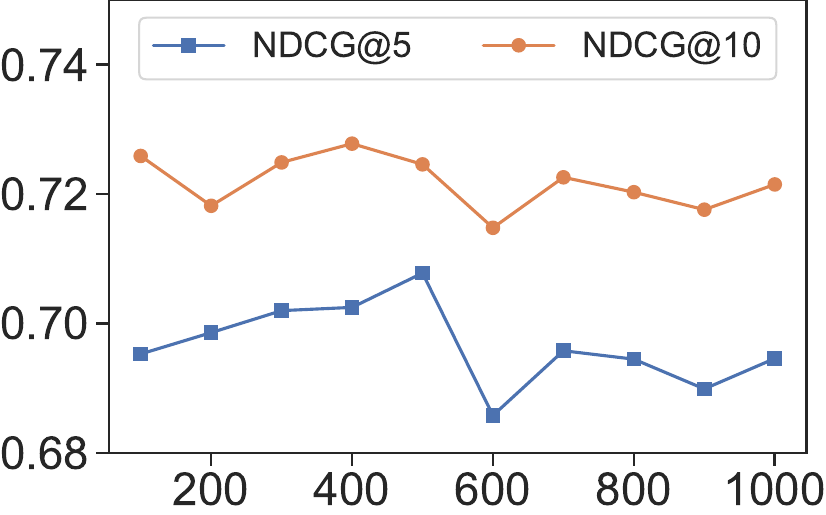}}
	\subfigure[Digital Art]{\includegraphics[width=0.32\columnwidth]{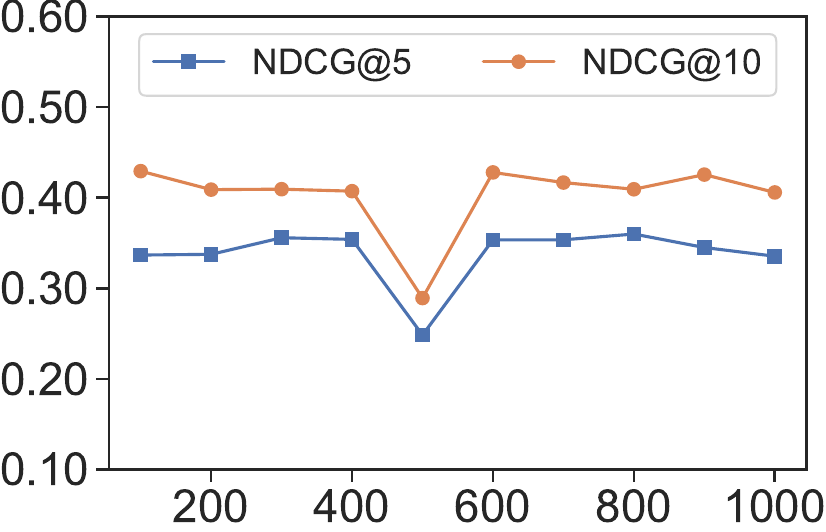}}
 	\subfigure[Rental]{\includegraphics[width=0.32\columnwidth]{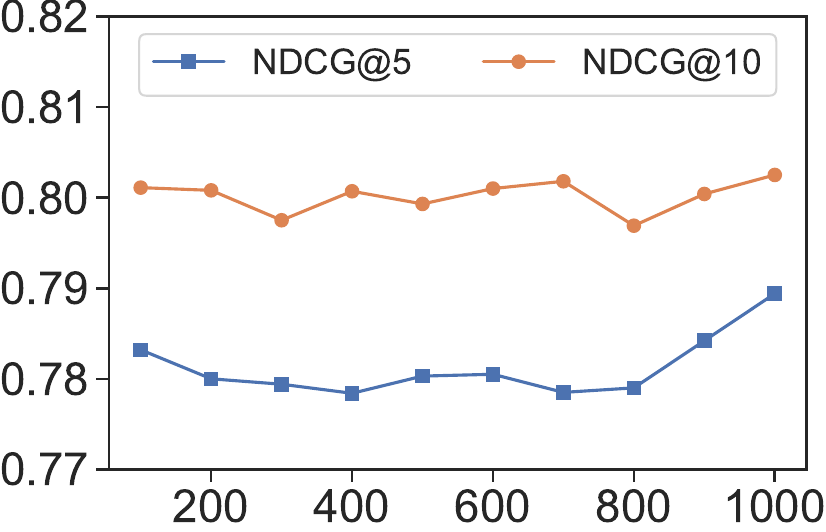}}
  
        \subfigure[Food Delivery]{\includegraphics[width=0.32\columnwidth]{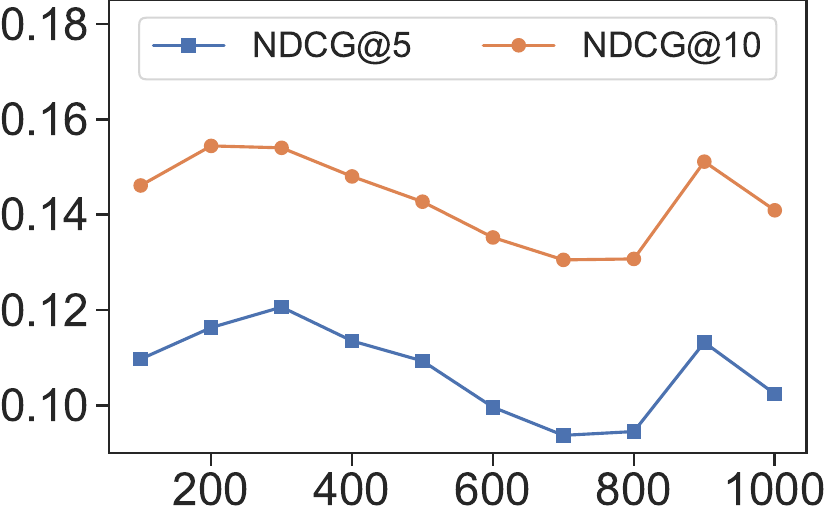}}
        \subfigure[Daily Necessity]{\includegraphics[width=0.32\columnwidth]{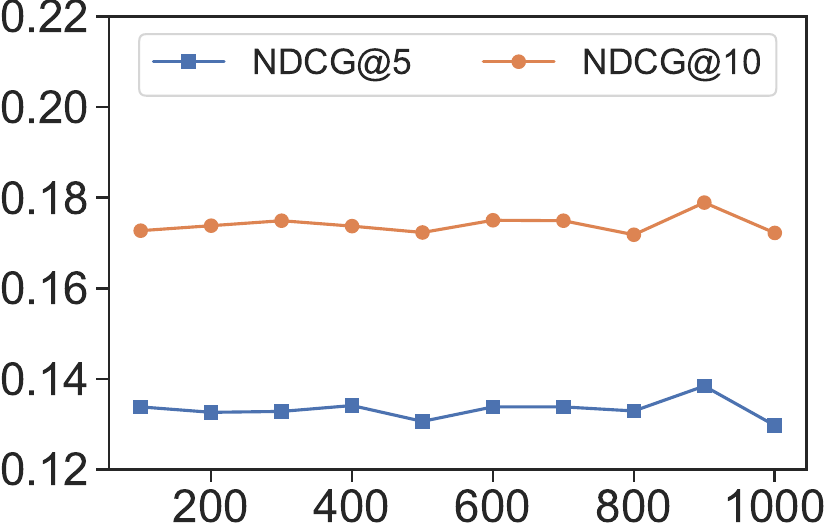}}
        \subfigure[Dining]{\includegraphics[width=0.32\columnwidth]{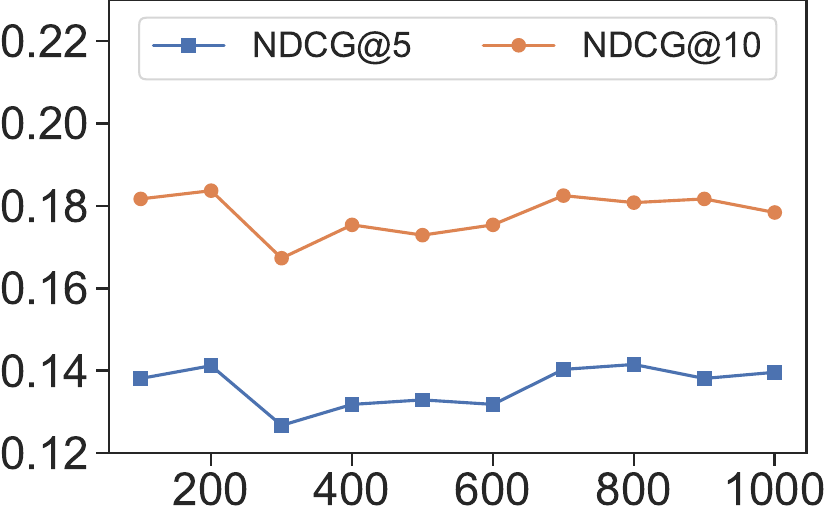}}
  \caption{ The influence of prototype number \wrt NDCG.}
	\label{fig:prototypes}
\end{figure}

Next, we 
investigate
the impact of the number of prototypes by varying it from 100 to 1000.
From 
Figure~\ref{fig:prototypes}, we 
observe 
{\model} always achieve superior performance with a large number of prototypes (\ie nearly (900, 1000)), which indicates that a large number of prototypes is a proper choice in real-world scenarios by 
i) better mitigating the noise in the huge entity graph,
and ii) endowing powerful capability for capturing complicated semantics.

\subsubsection{Zooming into the Prototype Enhanced Attention}
\begin{figure}
	\centering
	\subfigure[Travel]{\includegraphics[width=0.32\columnwidth]{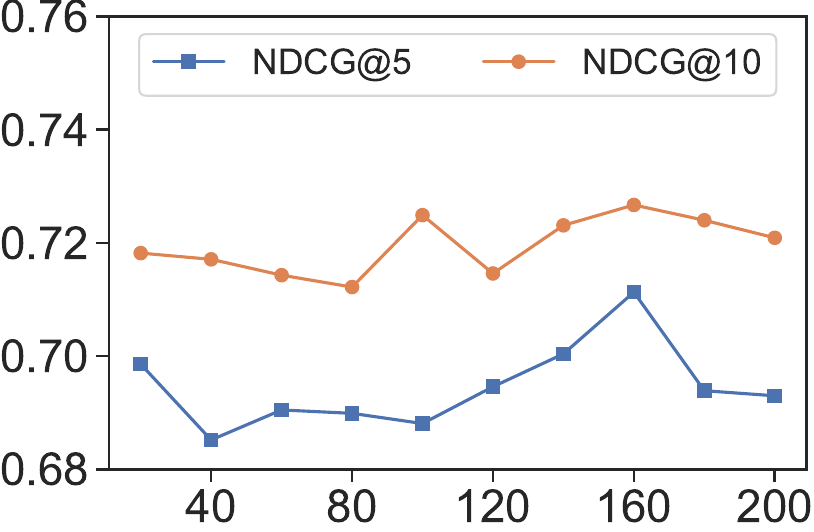}}
	\subfigure[Digital Art]{\includegraphics[width=0.32\columnwidth]{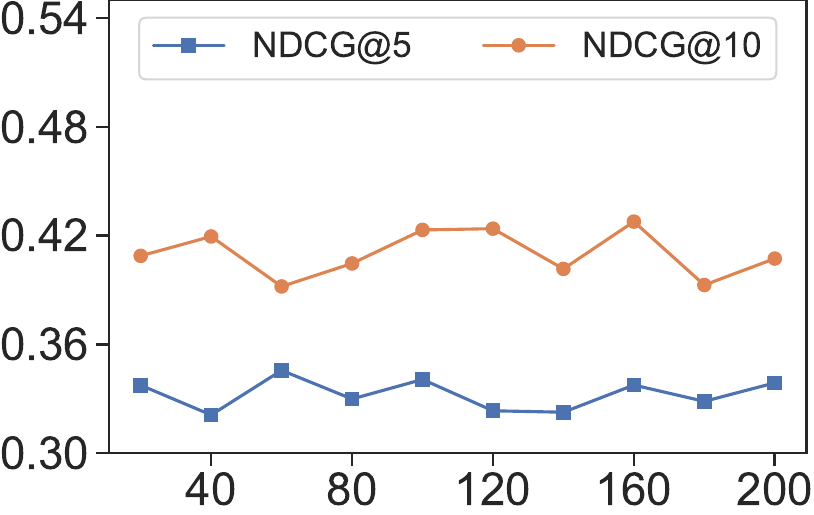}}
 	\subfigure[Rental]{\includegraphics[width=0.32\columnwidth]{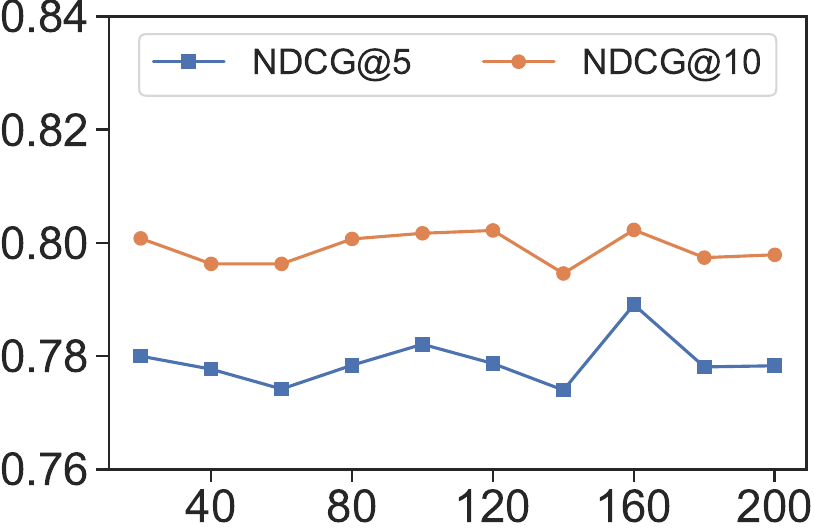}}
  
        \subfigure[Food Delivery]{\includegraphics[width=0.32\columnwidth]{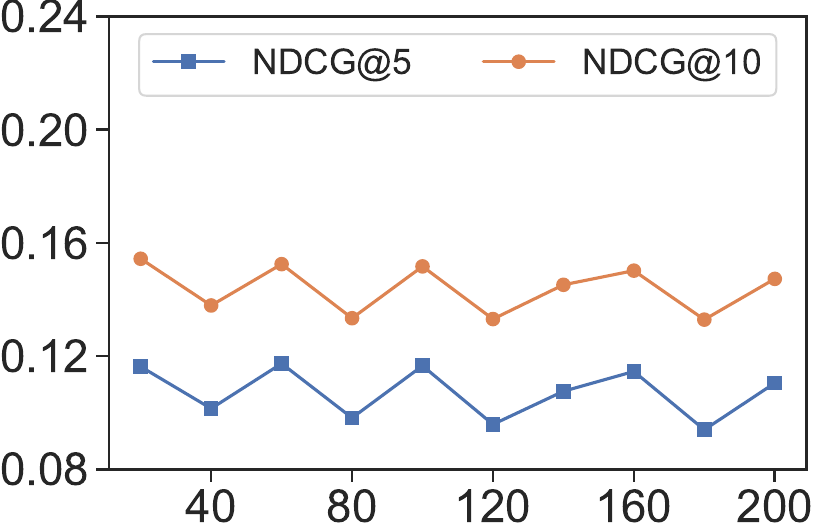}}
        \subfigure[Daily Necessity]{\includegraphics[width=0.32\columnwidth]{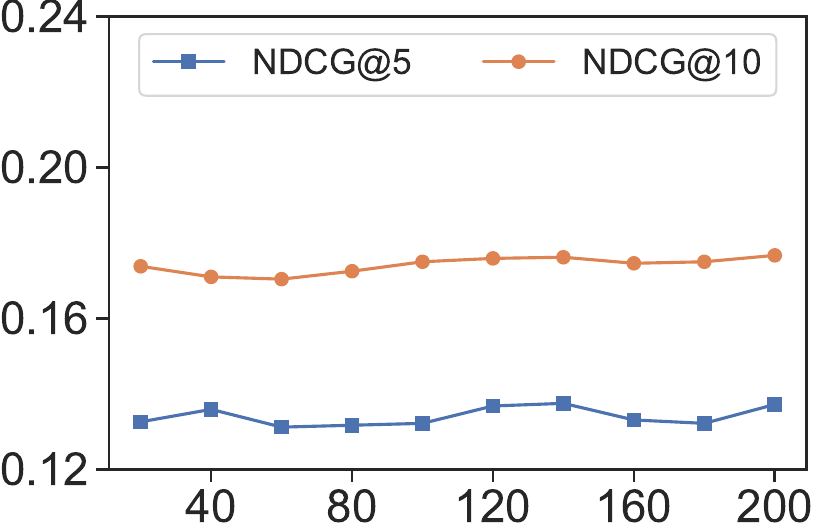}}
        \subfigure[Dining]{\includegraphics[width=0.32\columnwidth]{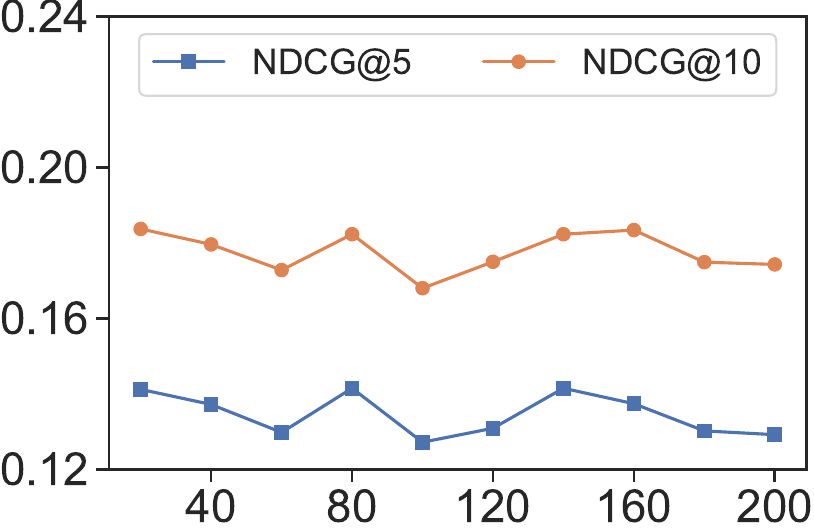}}
  \caption{ The study of the prototype enhanced attention  \wrt NDCG.}
	\label{fig:topk}
\end{figure}

In the prototype enhanced attention, we utilize a mask vector to sparsely activate the most relevant $K$ prototypes to characterize scenario-based context for user representations. 
Here, we investigate into the impact of the number of prototypes 
by varying it from 20 to 200. From the performance comparison results in Figure~\ref{fig:topk}, we observe that {\model} always achieves superior performance with $K$ in (100, 160), while too large or small $K$ would harm the model due to inadequate knowledge or unnecessary noises, respectively.

\subsection{Online Performance}
\begin{figure}[H]
	\centering
	\subfigure[Travel]{\includegraphics[width=0.31\columnwidth]{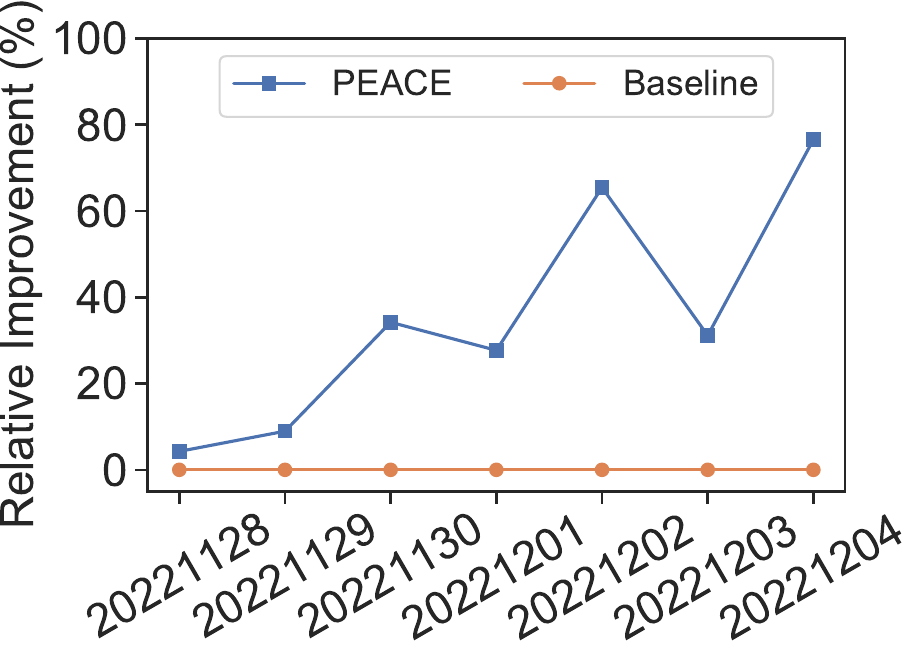}}
	\subfigure[Digital Art]{\includegraphics[width=0.31\columnwidth]{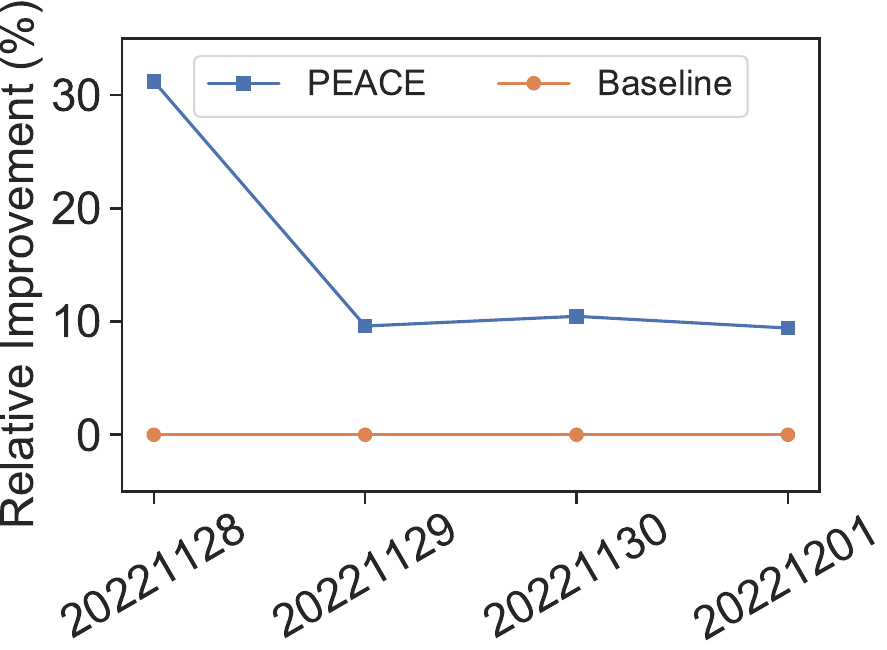}}
     \subfigure[Rental]{\includegraphics[width=0.31\columnwidth]{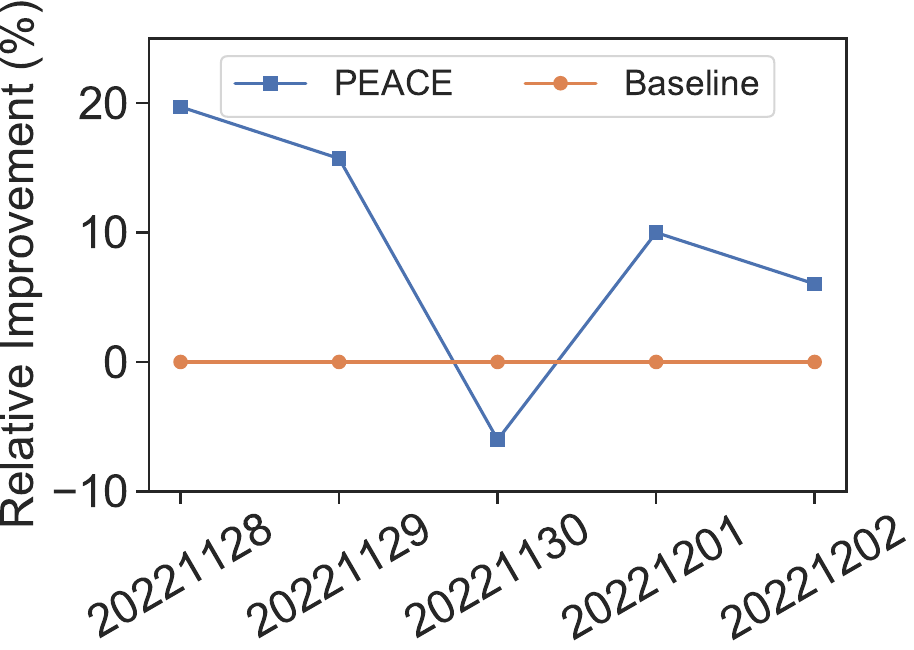}}
     
    \subfigure[Food Delivery]{\includegraphics[width=0.31\columnwidth]{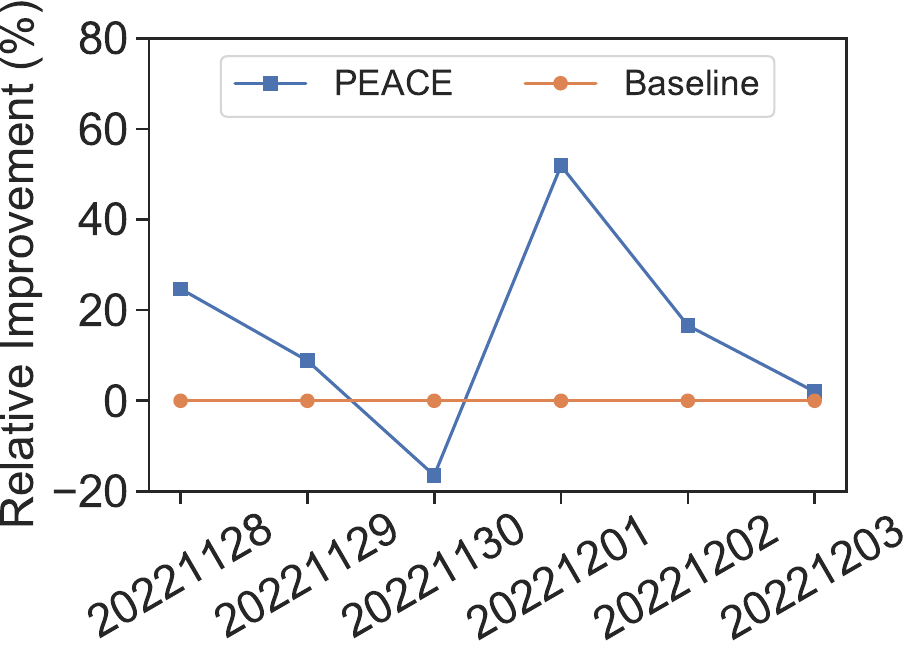}}
     \subfigure[Daily Necessity]{\includegraphics[width=0.31\columnwidth]{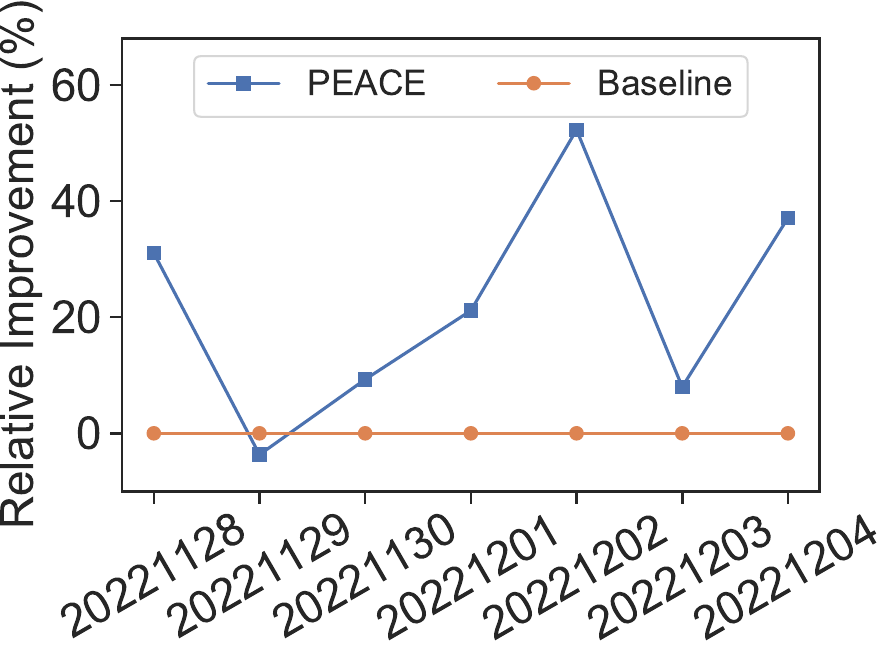}}
    	\subfigure[Dining]{\includegraphics[width=0.31\columnwidth]{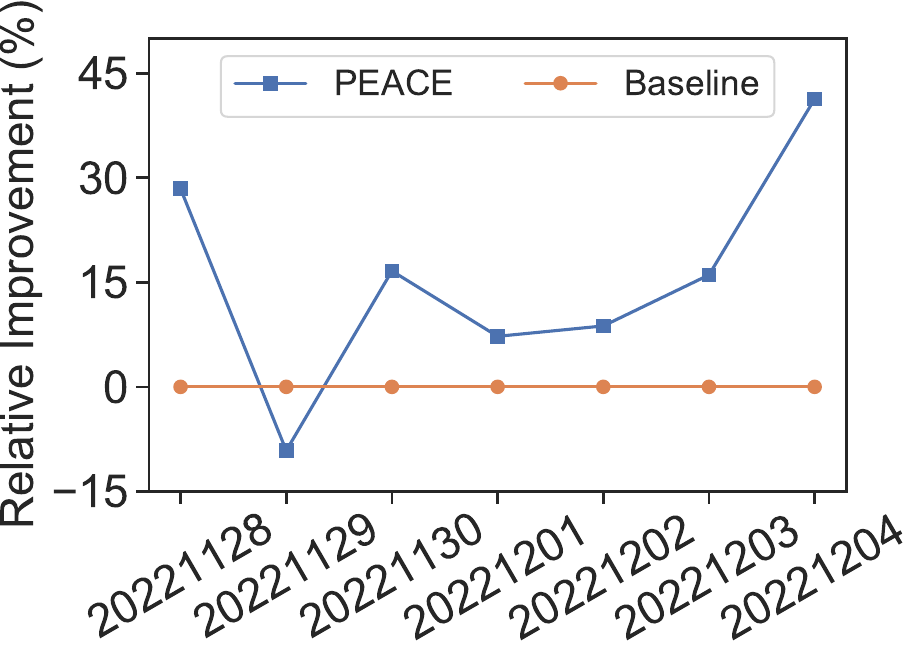}}
  
  \caption{Online performance on real-world scenarios.}
	\label{fig:online_ab}
\end{figure}

We
evaluate the proposed {\model} via real traffic in six downstream scenarios by competing against the deployed DeepFM-based baseline. Specifically, we perform the online evaluation from ``2022/11/28'' to `2022/12/04' with the widely-adopted CTR (\ie click-through rate) as the core metric, and report the experimental results in  Figure \ref{fig:online_ab}.
We find that {\model} always keeps the best performance across multiple downstream tasks, which is consistent with the offline evaluations. 
Overall
the proposed {\model} achieves 
statistically significant
performance gains 
by 35.63\%, 11.00\%, 4.13\%, 11.95\%, 15.87\% and 19.05\% 
with a significance level of 95\% 
in the six scenarios, respectively,
which further demonstrates the effectiveness of {\model} on cross-domain recommendation task in real-world service platforms by adopting a more principled way to distill universal and transferable knowledge for a variety of downstream applications. 

%% file: tab/data_tab.tex
\begin{table}[t]
    \centering
    \caption{Statistics of the source and target domains. }
    \label{tab:data_sta}
    \footnotesize
    \setlength{\tabcolsep}{4pt}
    \begin{tabular}{ c| c |c |c |c}
    \toprule
    \multirow{4}[5]{*}{\centering Source domains} & Datasets & \# Users & \# Entities & \# Interactions\\
    \cmidrule{2-5}
    {} & Payment & 452,626,664 & 5,866,019 & 5,300,961,103  \\
    \cmidrule{2-5}
    {} & Homepage Feed & 164,426,735 & 21,001 & 862,691,910  \\
    \cmidrule{2-5}
    {} & Search & 18,842,402 & 143,354 & 35,535,320   \\
    \midrule
     \multirow{7}[10]{*}{\centering Target domains} & Datasets & \# Users & \# Items & \# Interactions\\
     \cmidrule{2-5}
     {} &Travel & 252,981 & 261 & 1,547,821  \\
    \cmidrule{2-5}
    {} & Digital Art & 54,499 & 213 & 251,967  \\
    \cmidrule{2-5}
    {} & Rental  & 230,403 & 253 & 1,489,039  \\
    \cmidrule{2-5}
    {} &Food Delivery & 506,852 & 24,252 & 5,563,663  \\
    \cmidrule{2-5}
    {} & Daily Necessity & 450,391 & 22,396 & 4,921,769  \\
    \cmidrule{2-5}
    {} & Dining& 596,102 & 25,608 & 6,505,069  \\
    \midrule
    \multicolumn{2}{c|}{\centering \multirow{2}{*}{\centering Entity graph}}  & \# Entities & \# Relations & \# Triplets\\
    \cmidrule{3-5}
     \multicolumn{2}{c|}{} & {17,621,487} & {107} & {103,227,202} \\
    \bottomrule
    \end{tabular}        
\end{table}

%% file: tab/main_tab.tex
\begin{table*}[t]\centering
\setlength{\tabcolsep}{1mm}{

\caption{Overall performance evaluation on six target domains. The best results are highlighted in boldface.} 
\label{tab:overall_result2}
\begin{tabular}{c|c|c|c|c|c|c|c|c|c|c|c|c}

\toprule
\multirow{2}{*}{Methods}
& \multicolumn{4}{c|}{Travel} 
& \multicolumn{4}{c|}{Digital Art} 
& \multicolumn{4}{c}{Rental} \\    
\cmidrule{2-13}
 {} 
 & {Hit@5} & {Hit@10} & {NDCG@5} & {NDCG@10} 
 & {Hit@5} & {Hit@10} & {NDCG@5} & {NDCG@10} 
 & {Hit@5} & {Hit@10} & {NDCG@5} & {NDCG@10} \\
\midrule 
{DeepFM} 
& {0.4841} & {0.6148} & {0.6755} & {0.7079} 
& {0.3555} & {0.5547} & {0.2988} & {0.3648} 
& {0.5529} & {0.6713} & {0.7655} & {0.7930}\\
\midrule 
{SASRec} 
& {0.4845} & {0.6223} & {0.6751} & {0.7093} 
& {0.3503} & {0.5625} & {0.3051} & {0.3731} 
& {0.5949} & {0.6721} & {0.7826} & {0.7997}\\
{BERT4Rec} 
& {0.5126} & {0.6377} & {0.6882} & {0.7222} 
& {0.3659} & {0.6016} & {0.3311} & {0.4024} 
& {0.5787} & {0.6723} & {0.7777} & {0.8000}\\
{PeterRec} 
& {0.4768} & {0.6088} & {0.6802} & {0.7137} 
& {0.3451} & {0.5729} & {0.3020} & {0.3729} 
& {0.5757} & {\textbf{0.6733}} & {0.7733} & {0.7964}\\
{PAUP} 
& {0.4854} & {0.6187} & {0.6786} & {0.7125} 
& {0.3737} & {0.5938} & {0.3288} & {0.3913} 
& {0.5806} & {0.6685} & {0.7783} & {0.7988}\\
{ComiRec} 
& {0.4862} & {0.6168} & {0.6789} & {0.7121} 
& {0.3568} & {0.5807} & {0.3228} & {0.3893} 
& {0.5537} & {0.6423} & {0.7655} & {0.7868}\\
\midrule
{SimGCL} 
& {0.4909} & {0.5919} & {0.6813} & {0.7080} 
& {0.3138} & {0.5417} & {0.2624} & {0.3386}
& {0.5551} & {0.6385} & {0.7661} & {0.7859}\\
\midrule
{PCRec-L}
& {0.5108} & {0.5936} & {0.6829} & {0.7047}
& {0.3464} & {0.5677} & {0.3001} & {0.3682} 
& {0.5483} & {0.6709} & {0.7649} & {0.7918}\\
{PCRec-D} 
& {0.4978} & {0.5977} & {0.6793} & {0.7051}
& {0.3490} & {0.5690} & {0.3116} & {0.3802} 
& {0.5536} & {0.6699} & {0.7650} & {0.7919}\\
\midrule
{\model} 
& {\textbf{0.5378}} & {\textbf{0.6449}} & {\textbf{0.7025}} & {\textbf{0.7278}} 
& {\textbf{0.3854}} & {\textbf{0.7031}} & {\textbf{0.3392}} & {\textbf{0.4265}} 
& {\textbf{0.6082}} & {0.6709} & {\textbf{0.7894}} & {\textbf{0.8025}}\\
\midrule
\midrule
\multirow{2}{*}{Methods}
& \multicolumn{4}{c|}{Food Delivery} 
& \multicolumn{4}{c|}{Daily Necessity} 
& \multicolumn{4}{c}{Dining} \\    
\cmidrule{2-13}
 {} 
 & {Hit@5} & {Hit@10} & {NDCG@5} & {NDCG@10} 
 & {Hit@5} & {Hit@10} & {NDCG@5} & {NDCG@10} 
 & {Hit@5} & {Hit@10} & {NDCG@5} & {NDCG@10} \\
\midrule 
{DeepFM} 
& {0.0418} & {0.0717} & {0.0940} & {0.1265}
& {0.1305} & {0.2117} & {0.1276} & {0.1636} 
& {0.1401} & {0.2454} & {0.1244} & {0.1667}\\
\midrule
{SASRec} 
& {0.0449} & {0.0752} & {0.0997} & {0.1310} 
& {0.1336} & {0.2254} & {0.1130} & {0.1711} 
& {0.1432} & {0.2480} & {0.1259} & {0.1698}\\
{BERT4Rec} 
& {0.0486} & {0.0793} & {0.1098} & {0.1426} 
& {0.1318} & {0.2200} & {0.1267} & {0.1660} 
& {0.1455} & {0.2588} & {0.1290} & {0.1747}\\
{PeterRec} 
& {0.0465} & {0.0758} & {0.1036} & {0.1348} 
& {0.1364} & {\textbf{0.2315}} & {0.1314} & {0.1740} 
& {0.1410} & {0.2451} & {0.1254} & {0.1676}\\
{PAUP} 
& {0.0421} & {0.0778} & {0.0917} & {0.1303} 
& {0.1390} & {0.2253} & {0.1344} & {0.1727} 
& {0.1523} & {0.2568} & {0.1310} & {0.1739}\\
{ComiRec} 
& {0.0478} & {0.0775} & {0.1118} & {0.1432} 
& {0.1372} & {0.2243} & {0.1324} & {0.1711} 
& {0.1445} & {0.2562} & {0.1262} & {0.1720}\\
\midrule
{SimGCL} 
& {0.0457} & {0.0742} & {0.0992} & {0.1291} 
& {0.1359} & {0.2216} & {0.1330} & {0.1713} 
& {0.1491} & {0.2497} & {0.1326} & {0.1729}\\
\midrule
{PCRec-L}
& {0.0438} & {0.0739} & {0.0987} & {0.1311} 
& {0.1340} & {0.2178} & {0.1303} & {0.1679} 
& {0.1348} & {0.2389} & {0.1173} & {0.1597}\\
{PCRec-D} 
& {0.0469} & {0.0754} & {0.1086} & {0.1389} 
& {0.1306} & {0.2142} & {0.1286} & {0.1659} 
& {0.1423} & {0.2500} & {0.1250} & {0.1689}\\
\midrule
{\model} 
& {\textbf{0.0503}} & {\textbf{0.0824}} & {\textbf{0.1206}} & {\textbf{0.1540}} 
& {\textbf{0.1397}} & {0.2290} & {\textbf{0.1384}} & {\textbf{0.1789}} 
& {\textbf{0.1618}} & {\textbf{0.2679}} & {\textbf{0.1412}} & {\textbf{0.1837}}\\
\bottomrule
\end{tabular}}

\end{table*}

%% file: tab/main_table_zero_shot.tex
\begin{table}[t]\centering
\caption{Performance comparison of zero-shot recommendation across six target domains. }
\setlength{\tabcolsep}{1mm}{
\begin{tabular}{c|c|c|c|c|c}
\toprule
\multirow{2}{*}{Domains}
& \multirow{2}{*}{Methods} 
& \multicolumn{4}{c}{Metrics} \\   
\cline{3-6}
 {} & {} & {Hit@5} & {Hit@10} & {NDCG@5} & {NDCG@10}   \\
\hline 
\multirow{2}{*}{Travel}
& {Tiger} & {0.0130} & {0.0181} & {0.0100} & {0.0123}  \\
& {\model} & {\textbf{0.5216}} & {\textbf{0.5968}} & {\textbf{0.6905}} & {\textbf{0.7104}}  \\ 
\cline{1-6}
{Digital}
& {Tiger} & {0.2344} & {\textbf{0.3516}} & {\textbf{0.2035}} & {\textbf{0.2561}}  \\
{Art} & {\model} & {\textbf{0.2382}} & {\textbf{0.3516}} & {0.1763} & {0.2100}  \\  \cline{1-6}
\multirow{2}{*}{Rental}
& {Tiger} & {0.0141} & {0.0186} & {0.0104} & {0.0124}  \\
& {\model} & {\textbf{0.0171}} & {\textbf{0.0432}} & {\textbf{0.0153}} & {\textbf{0.0271}}  \\ 
\cline{1-6}
{Food}
& {Tiger} & {0.0200} & {0.0321} & {0.0524} & {0.0660}  \\
{Delivery}& {\model} & {\textbf{0.0314}} & {\textbf{0.0557}} & {\textbf{0.0765}} & {\textbf{0.1036}}  \\
\cline{1-6}
{Daily}
& {Tiger} & {0.0173} & {0.0349} & {0.0479} & {0.0718}  \\
{Necessity}& {\model} & {\textbf{0.0762}} & {\textbf{0.1608}} & {\textbf{0.0637}} & {\textbf{0.1002}}  \\ 
\cline{1-6}
\multirow{2}{*}{Dining}
& {Tiger} & {0.0142} & {0.0290} & {0.0390} & {0.0597}  \\
& {\model} & {\textbf{0.1491}} & {\textbf{0.2503}} & {\textbf{0.1324}} & {\textbf{0.1735}}  \\
\bottomrule
\end{tabular}}

\label{tab:zero_shot_result}
\end{table}

%% file: sec-con.tex
\section{Conclusion}
In this paper, we propose {\model}, a prototype learning augmented transferable framework for cross-domain
recommendation, which is built upon a multi-interest and entity-oriented pre-training architecture. By incorporating the prototype learning, representations of users and items are greatly improved by the contrastive prototype learning module and the prototype enhanced attention mechanism. 
Extensive experiments on real-world datasets 
demonstrate that {\model} significantly outperforms state-of-the-art baselines in various scenarios with both offline and online settings.

%% file: sec-app.tex
\appendix
\section{Supplementary Material}
\subsection{Environment}
Due to the large volume of data in our experiments, all models are implemented on parameter server based distributed learning systems. Specifically, we utilize 20 worker with 12 cores and 100GB memory, and 20 parameter servers with 5 cores and 15GB memory for the training of all models.

\subsection{Evaluation Protocol}
We mainly evaluate the proposed {\model} for transferable recommendation in the normal and zero-shot settings. After the pre-training stage, the proposed {\model} is evaluated as follows:
\begin{itemize}[leftmargin=*]
    \item \textbf{Normal recommendation}: For each target domain, we do a chronological train/validation/test split with 70\%/15\%/15\% according to interaction timestamps.
    \item \textbf{Zero-shot recommendation}: For each target domain, we directly apply the pre-trained {\model} for prediction, \ie all interactions in the domain are used for test.
\end{itemize}

\subsection{Parameter Settings}
For all methods, we set the batch size as 512, embedding size as 64 and adopt Adam as the optimizer with a learning rate of 1e-4. 
Moreover, we set the number of GNN layers as 2 for all GNN-based methods and the sequence length as 200 for all sequential methods. 
All MLP($\cdot$) components are commonly implemented with two hidden layers with ReLU as the activation function. 
It is worthwhile that these sequential learning methods (\ie SASRec, BERTRec,PeterRec, PAUP, ComiRec) cannot directly applied to our task since items in source and target domains are completely non-overlapping. \textbf{Therefore, we utilize the entity-oriented pre-training strategy in these approaches, \ie replacing the original item-wise sequences with entity-wise sequences through our item-entity mapping function.}
For a fair comparison, we also utilize the widely-adopted grid search strategy to carefully tune each approach in the validation set
and briefly present the optimal parameters as follows:

\paratitle{For baselines}
\begin{itemize}[leftmargin=*]
    \item \textbf{DeepFM}: the network structure is 64-32-8 and the latent dimension of FM is 8. To be fair, in the fine-tuning stage, all pre-trained models are under the same setting.
    \item \textbf{SASRec}: the number of self-attention blocks is set to 2 and the learned positional embedding is utilized.
    \item \textbf{BERT4Rec}: the layer number $L$ is set to 2, the number of heads $h$ is set to 2 and the learnable positional embedding is used.
    \item \textbf{PeterRec}: the kernel size, the number of dilation channels and dilations are set to 3, 64 and $\left [ 1,4,1,4,1,4,1,4 \right ]$ respectively.
    \item \textbf{PAUP}: the layer $L$, head number $h$ and segment number $m$ are set to 2, 2 and 5 respectively, and the unsymmetrical positional encoding is used.
    \item \textbf{ComiRec}: the number of interest embeddings is set to 8.
    \item \textbf{SimGCL}: the parameter for $L_2$ regularization and temperature $\tau$ are set to 1e-4 and 0.2 respectively.
    \item \textbf{PCRec-L}: the parameter for $L_2$ regularization is set to 1e-4 and layer combination coefficient $\alpha_k$ is uniformly set to $\frac{1}{3}$ given the predefined GNN layer number is 2.
    \item \textbf{PCRec-D}: the factor number is set to 8 with 2-layer GNN.
    \item \textbf{Tiger}: the number of discarded C is set to 1 with a 2-layer GNN.
\end{itemize}

\paratitle{For {\model}}, we empirically set the
temperature $\tau$  as 0.2. As for $\gamma$, the interest kernel number $M$, the number of prototypes $N$ and prototype enhanced attention parameter $K$, we set them to 1.0, 10, 900 and 160 according to the grid search results.

\subsection{Additional Results}
In this section, we provide the additional results of the influence of the entity graph, the influence of multi-interest encoding, and performance comparison \wrt the Hit metric and finally the efficiency analysis of the proposed {\model}.

\begin{figure}[h]
\vspace{-1.25em}
	\centering
	\subfigure[Travel]{\includegraphics[width=0.32\columnwidth]{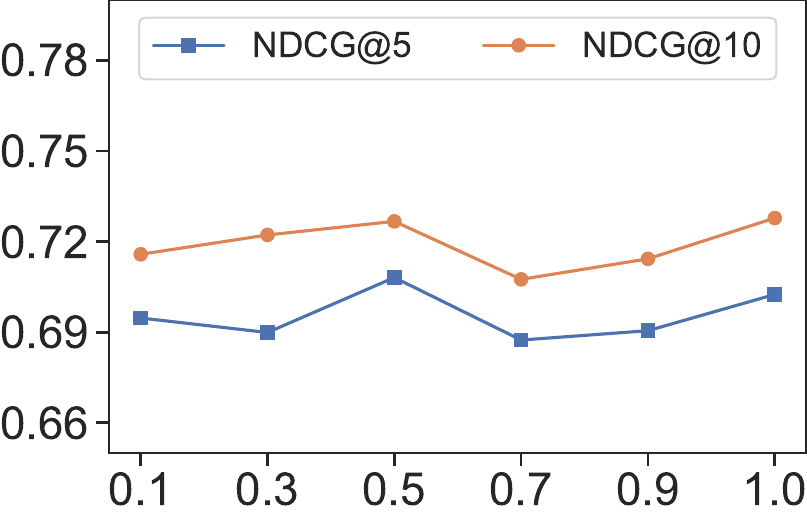}}
	\subfigure[Digital Art]{\includegraphics[width=0.32\columnwidth]{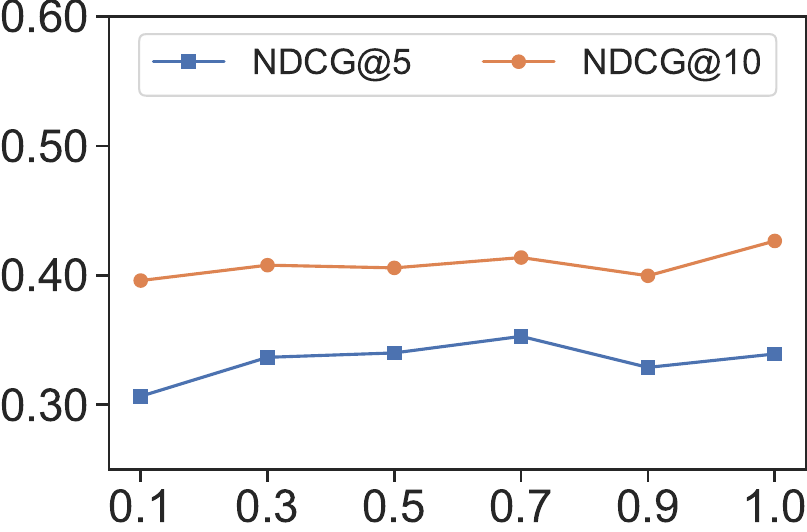}}
 	\subfigure[Rental]{\includegraphics[width=0.32\columnwidth]{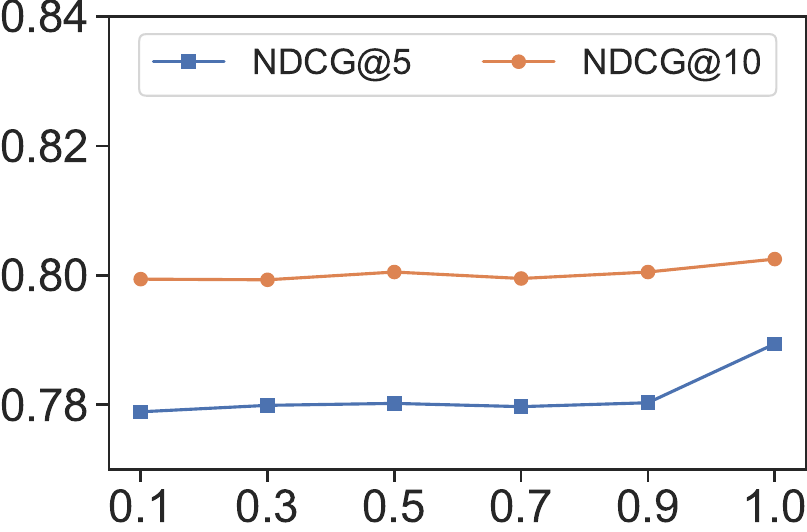}}
  
        \subfigure[Food Delivery]{\includegraphics[width=0.32\columnwidth]{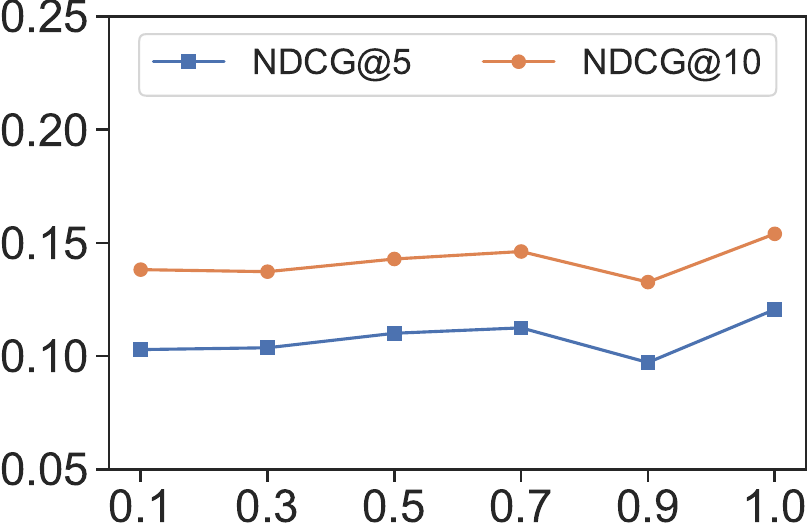}}
        \subfigure[Daily Necessity]{\includegraphics[width=0.32\columnwidth]{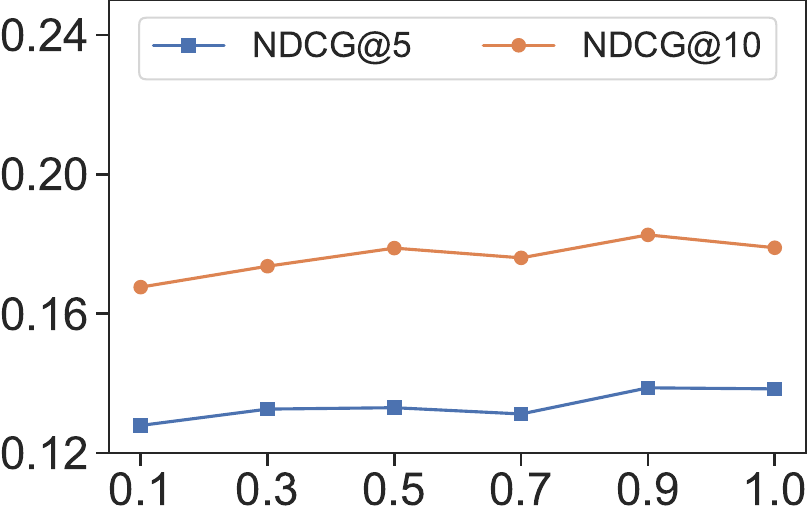}}
        \subfigure[Dining]{\includegraphics[width=0.32\columnwidth]{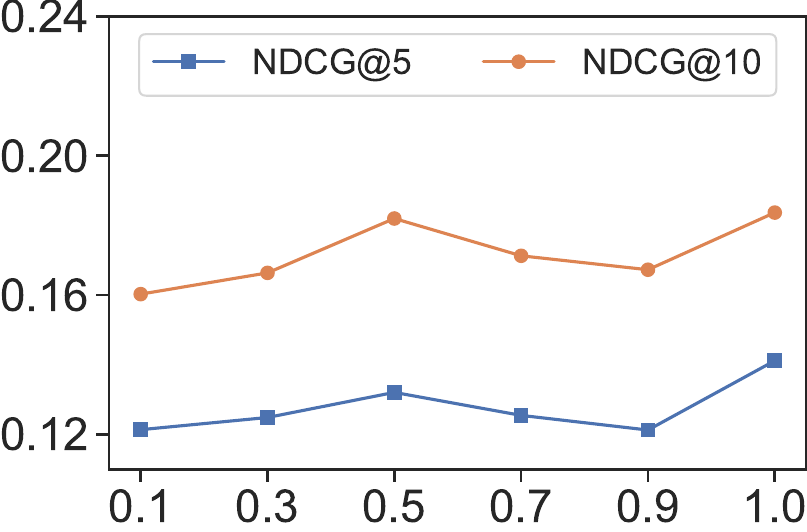}}
  \caption{The influence of entity graph \wrt NDCG.}
	\label{fig:kg}
	\vspace{-1.25em}
\end{figure}

\begin{figure}[h]
\vspace{-1.25em}
	\centering
	\subfigure[Travel]{\includegraphics[width=0.32\columnwidth]{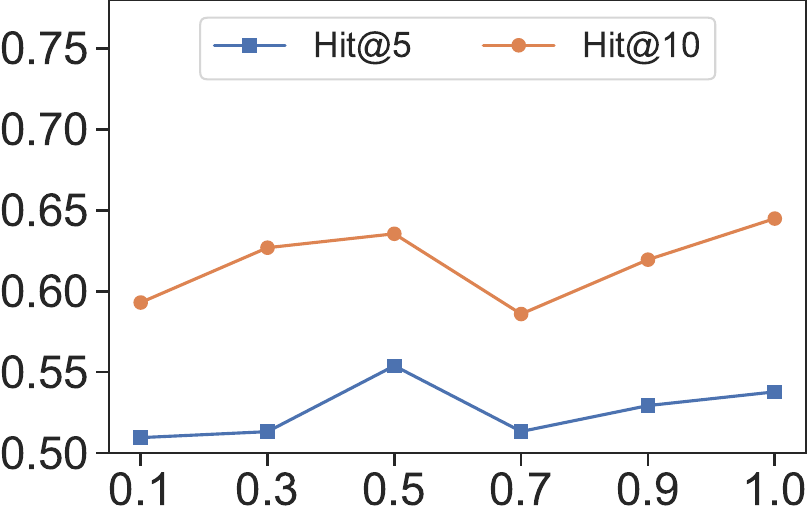}}
	\subfigure[Digital Art]{\includegraphics[width=0.32\columnwidth]{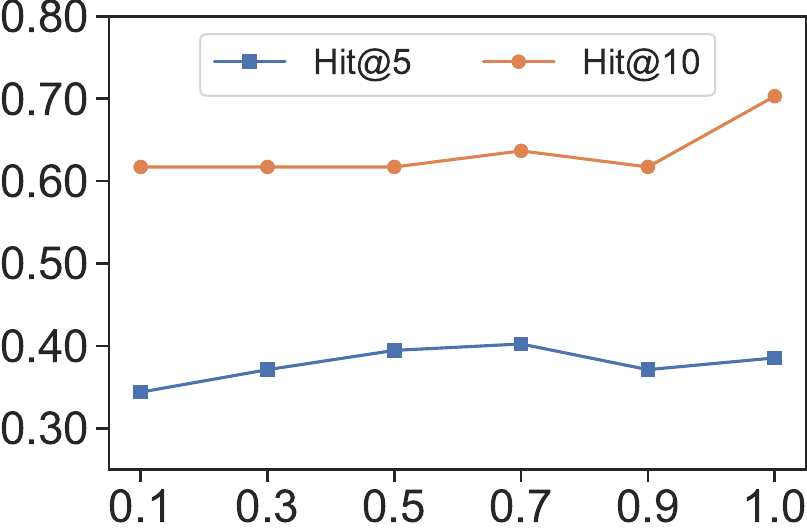}}
 	\subfigure[Rental]{\includegraphics[width=0.32\columnwidth]{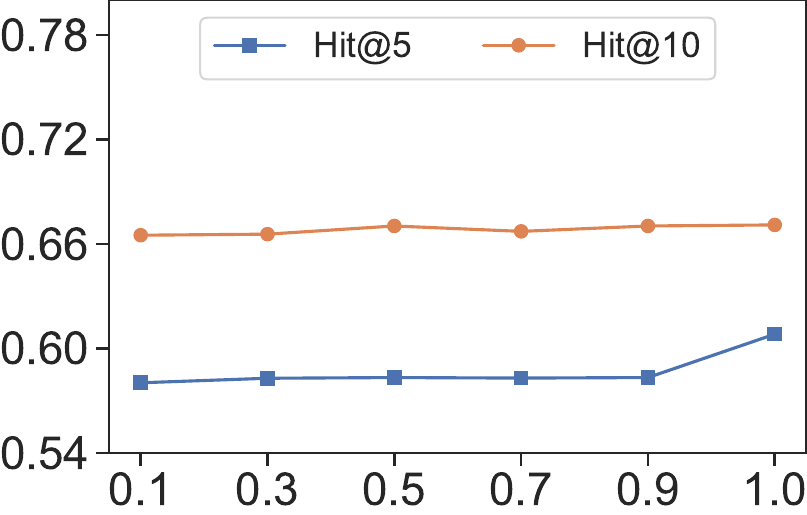}}
  
        \subfigure[Food Delivery]{\includegraphics[width=0.32\columnwidth]{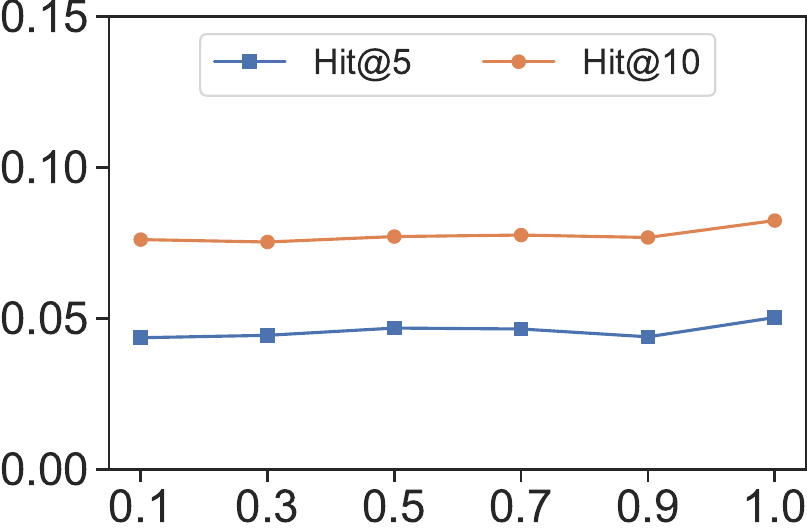}}
        \subfigure[Daily Necessity]{\includegraphics[width=0.32\columnwidth]{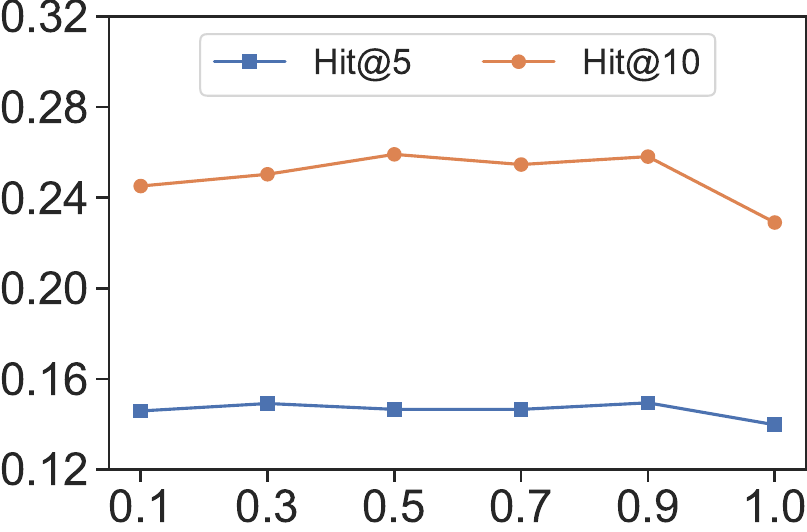}}
        \subfigure[Dining]{\includegraphics[width=0.32\columnwidth]{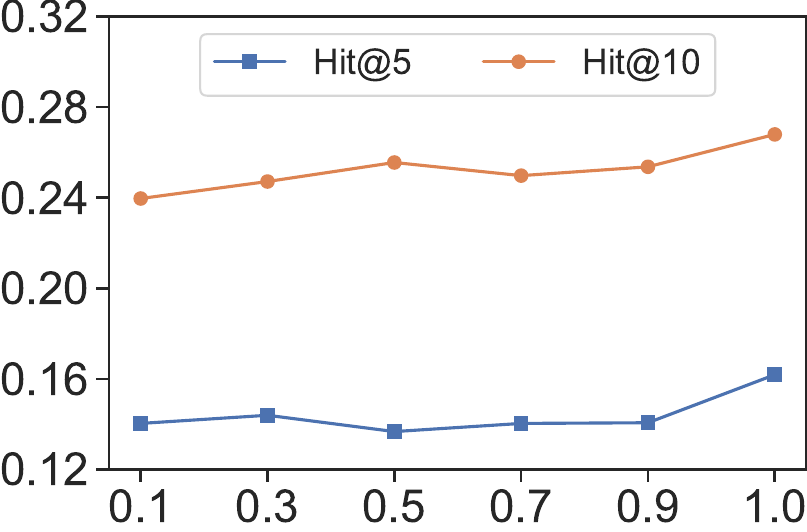}}
  \caption{The influence of entity graph \wrt Hit.}
	\label{fig:kg2}
	\vspace{-1.25em}
\end{figure}

\subsubsection{Influence of the Entity Graph}
Intuitively, the entity graph serves as a critical bridge for broadening user interests and discovering correlations between items. Hence, we measure how {\model} perform with different sparsity of the entity graph. 
Concretely, we randomly drop entity-entity relations to vary the sparsity of the entity graph in \{0.1, 0.2, 0.5, 0.7, 0.9, 1.0\}, and present the performance comparison in Figure \ref{fig:kg} and Figure \ref{fig:kg2}.
From the plot we find that the performance of {\model} generally achieves positive gains with the growth of the entity graph, which demonstrates the usefulness of the graph data, worthy of being comprehensively exploited in industrial recommenders.

\begin{figure}[h]
\vspace{-1.25em}
	\centering
	\subfigure[Travel]{\includegraphics[width=0.32\columnwidth]{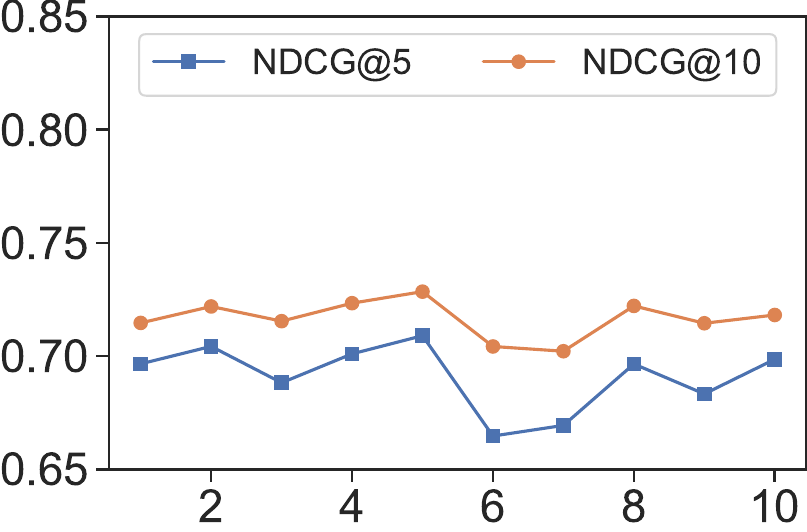}}
	\subfigure[Digital Art]{\includegraphics[width=0.32\columnwidth]{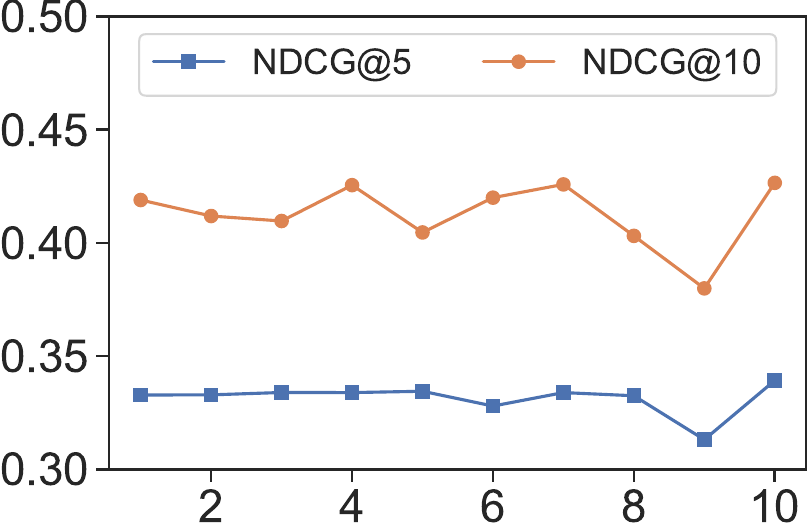}}
 	\subfigure[Rental]{\includegraphics[width=0.32\columnwidth]{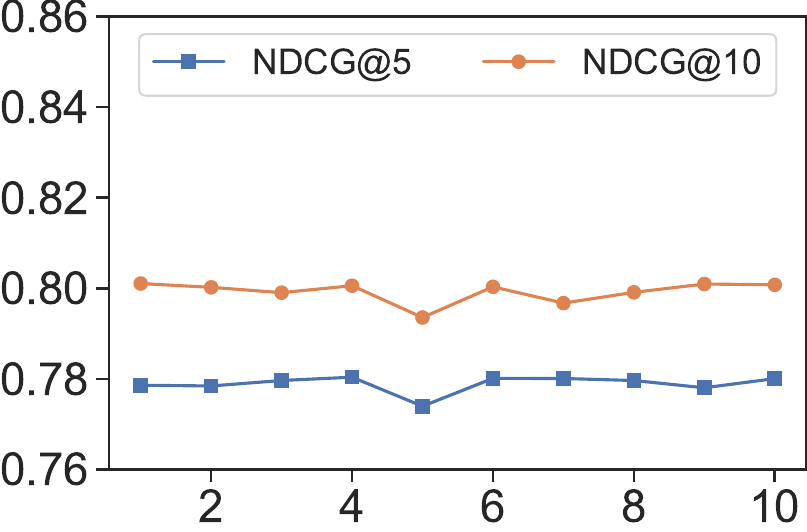}}
  
        \subfigure[Food Delivery]{\includegraphics[width=0.32\columnwidth]{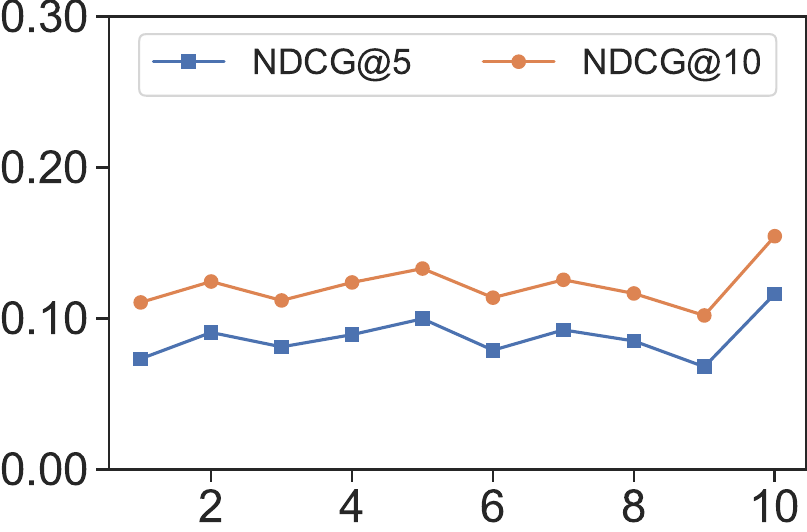}}
        \subfigure[Daily Necessity]{\includegraphics[width=0.32\columnwidth]{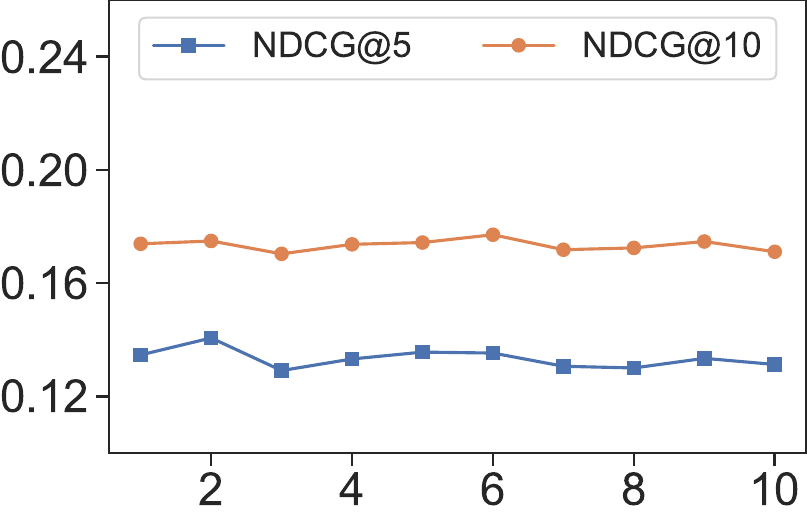}}
        \subfigure[Dining]{\includegraphics[width=0.32\columnwidth]{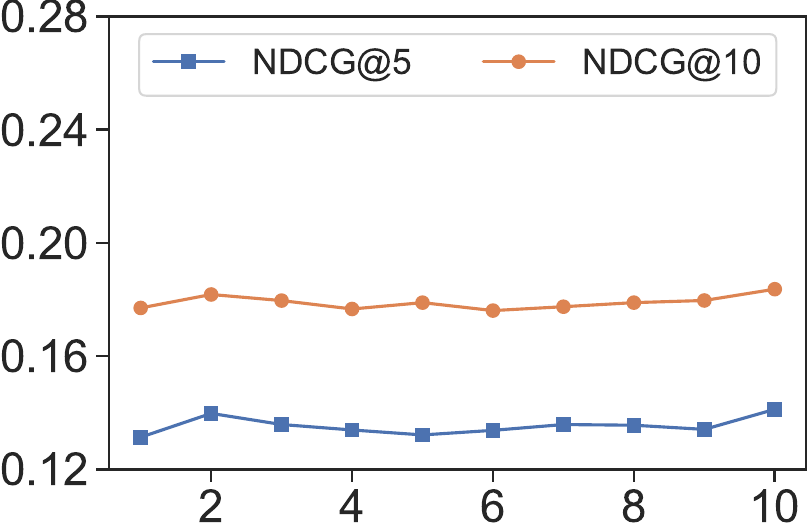}}
  \caption{ The study of multi-interest encoding \wrt NDCG.}
	\label{fig:multi_interest_2}
\end{figure}

\begin{figure}[h]
\vspace{-1.25em}
	\centering
	\subfigure[Travel]{\includegraphics[width=0.32\columnwidth]{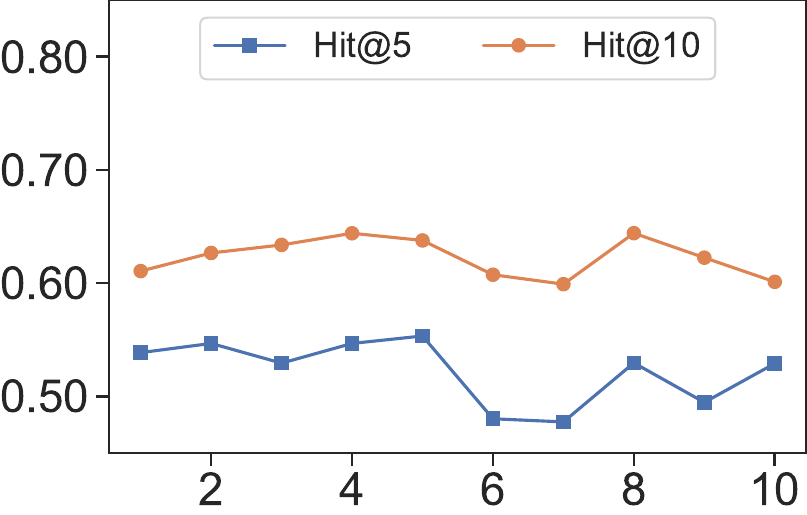}}
	\subfigure[Digital Art]{\includegraphics[width=0.32\columnwidth]{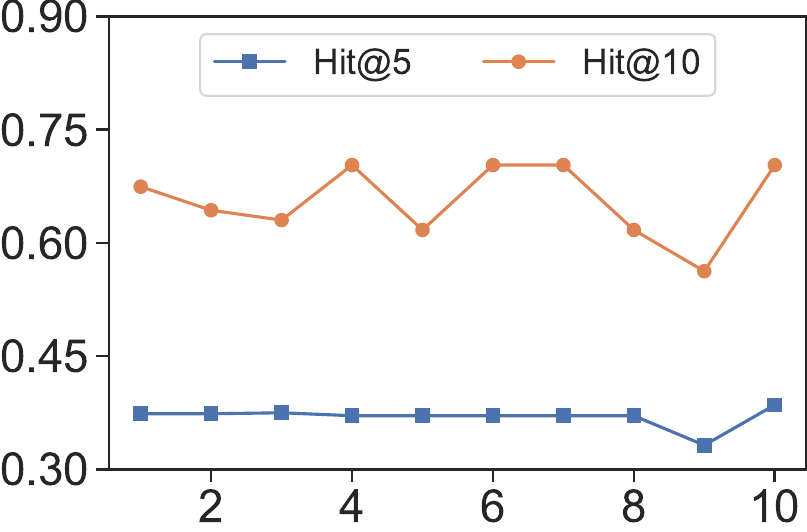}}
 	\subfigure[Rental]{\includegraphics[width=0.32\columnwidth]{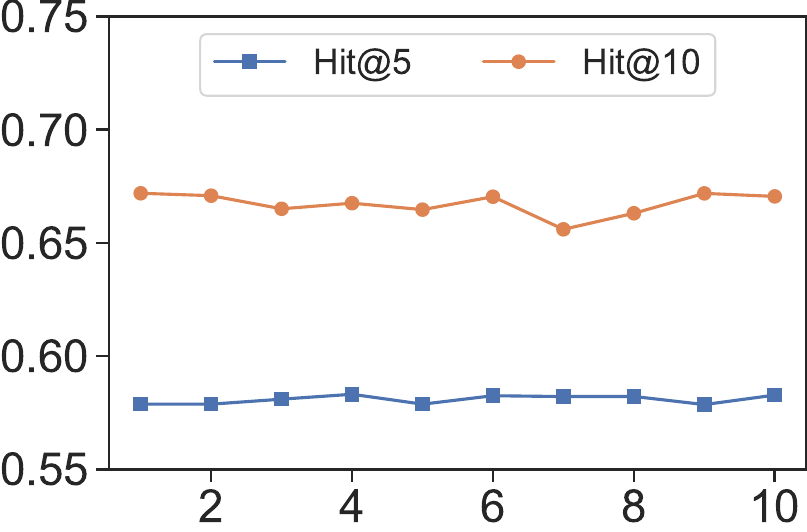}}
  
        \subfigure[Food Delivery]{\includegraphics[width=0.32\columnwidth]{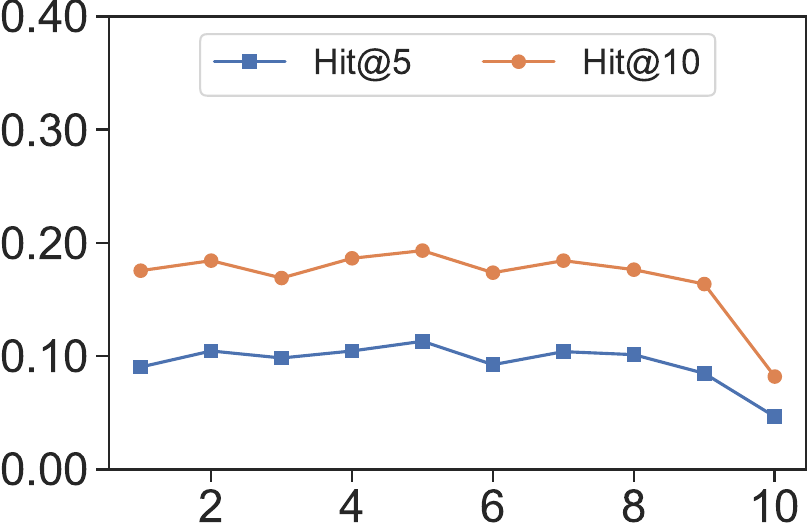}}
        \subfigure[Daily Necessity]{\includegraphics[width=0.32\columnwidth]{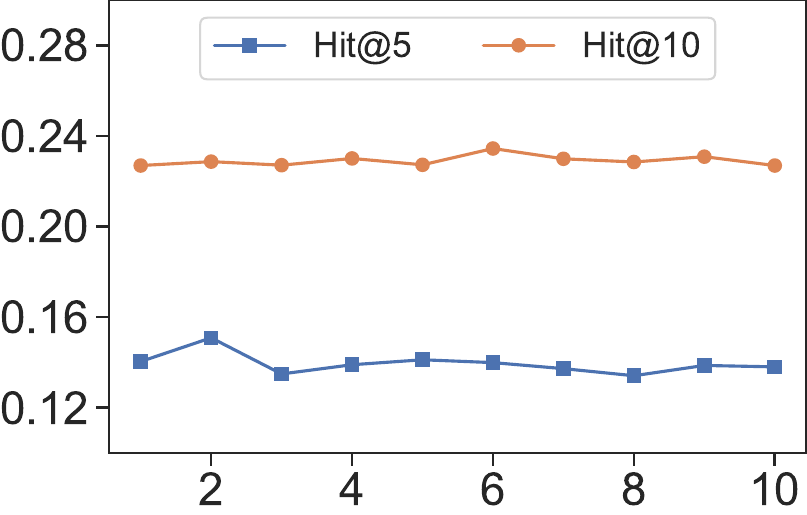}}
        \subfigure[Dining]{\includegraphics[width=0.32\columnwidth]{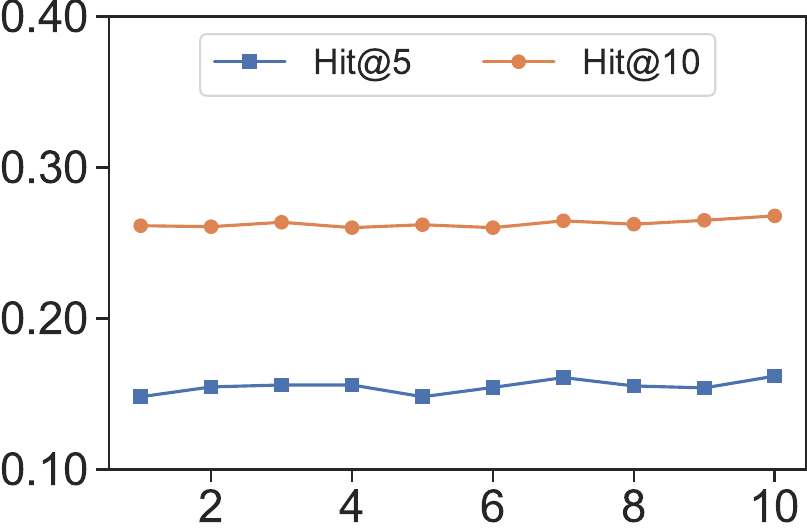}}
  \caption{ The study of multi-interest encoding \wrt Hit.}
	\label{fig:multi_interest_1}
\end{figure}

\subsubsection{Influence of multi-interest encoding}
In fact, user interests in multiple domains are quite complex and the multi-interest encoding can serve as a useful tool for describing the complex user interests.
Concretely, we investigate into the influence of multi-interest encoding by varying the interest kernels $M$ from 1 to 10 and present the performance comparison in Figure~\ref{fig:multi_interest_2} and Figure~\ref{fig:multi_interest_1}.

From Figure~\ref{fig:multi_interest_2} and Figure~\ref{fig:multi_interest_1} we find that the performance of {\model} generally improves as the number of interests grows, which demonstrates that the fine-grained characterization of user interests are helpful due to the multiplex data in real-world service platforms.

\subsubsection{Performance comparison results \wrt the Hit metric}
\begin{figure}[h]
\vspace{-1.25em}
	\centering
	\subfigure[Hit@5]{\includegraphics[width=0.45\columnwidth]{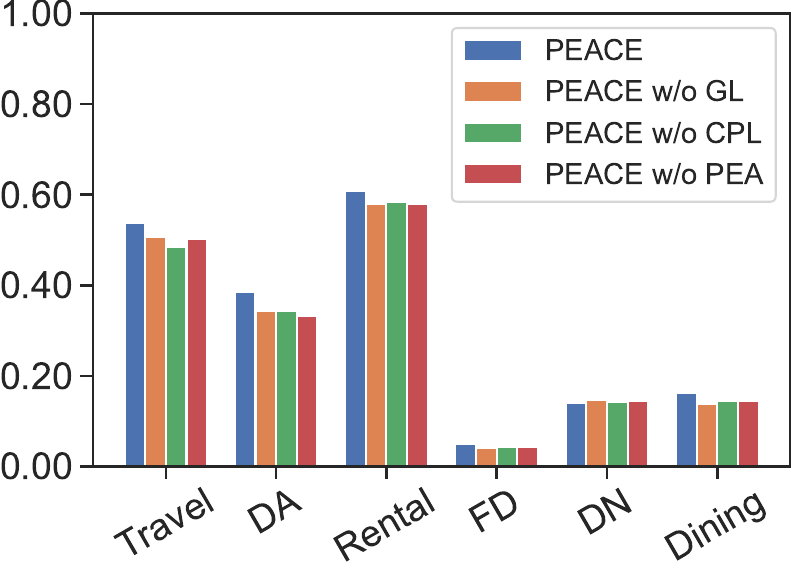}}
	\subfigure[Hit@10]{\includegraphics[width=0.45\columnwidth]{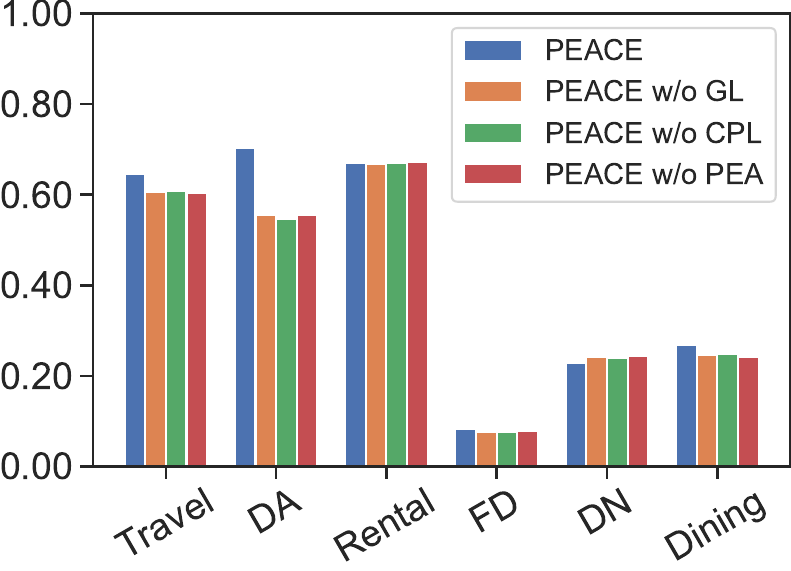}}
  \caption{Ablation study of {\model} \wrt Hit.}
	\label{fig:ab_study2}
\end{figure}

\begin{figure}[h]
\vspace{-1.25em}
	\centering
	\subfigure[Travel]{\includegraphics[width=0.32\columnwidth]{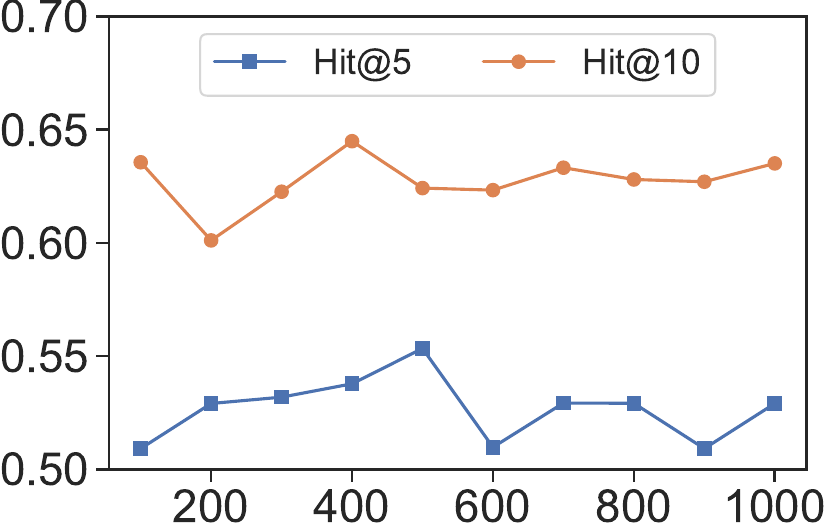}}
	\subfigure[Digital Art]{\includegraphics[width=0.32\columnwidth]{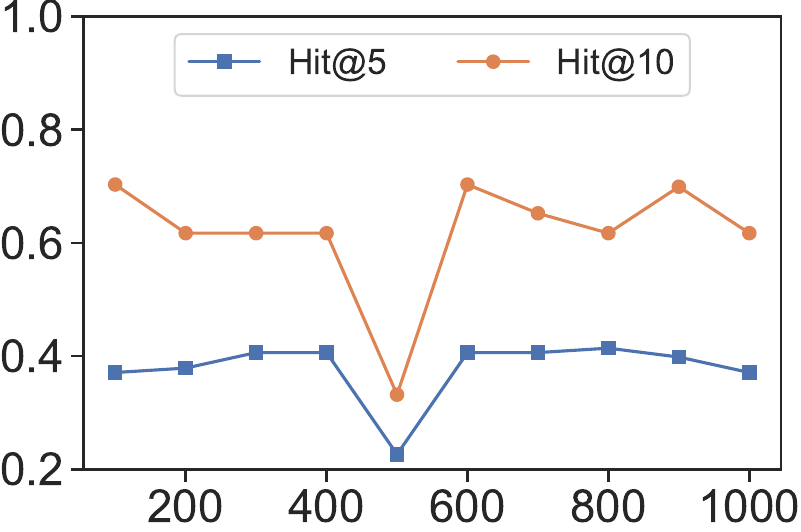}}
 	\subfigure[Rental]{\includegraphics[width=0.32\columnwidth]{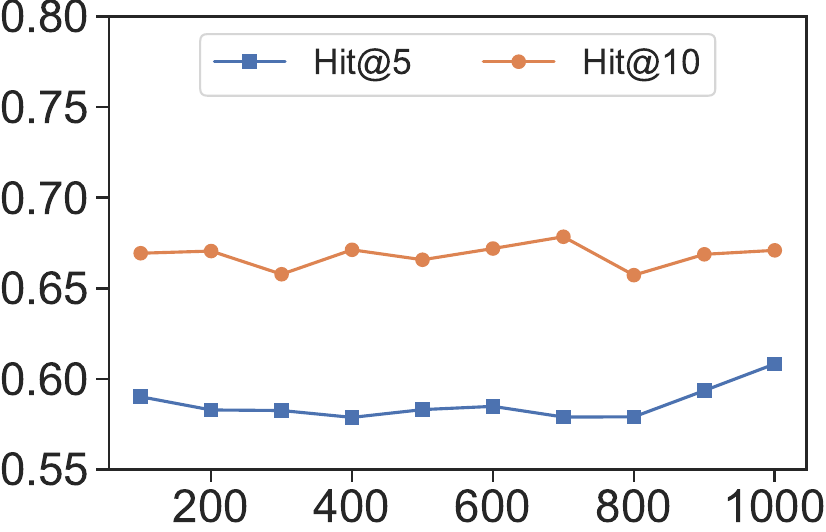}}
  
        \subfigure[Food Delivery]{\includegraphics[width=0.32\columnwidth]{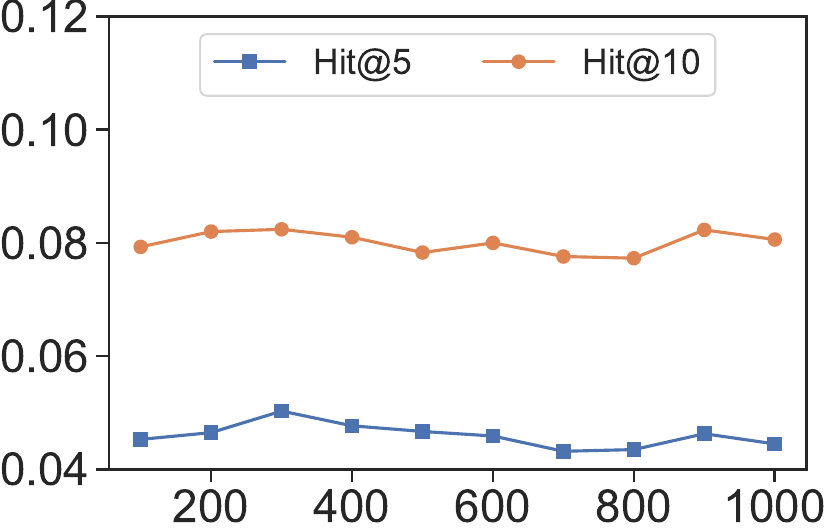}}
        \subfigure[Daily Necessity]{\includegraphics[width=0.32\columnwidth]{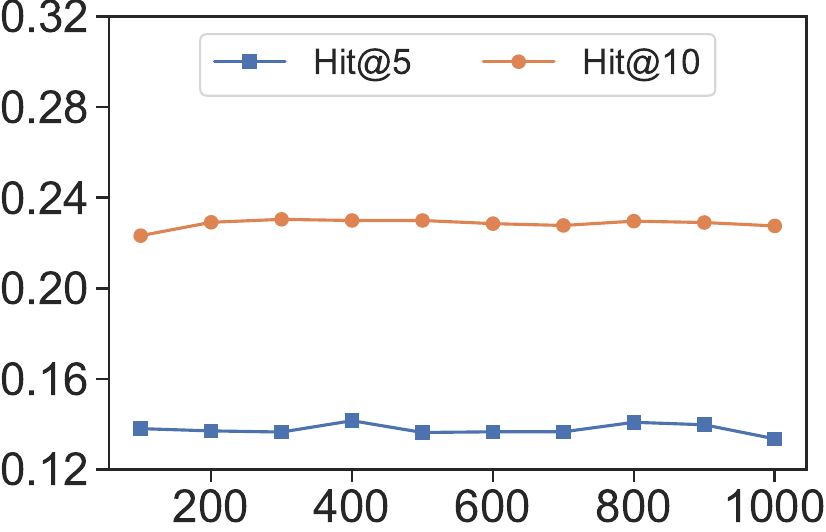}}
        \subfigure[Dining]{\includegraphics[width=0.32\columnwidth]{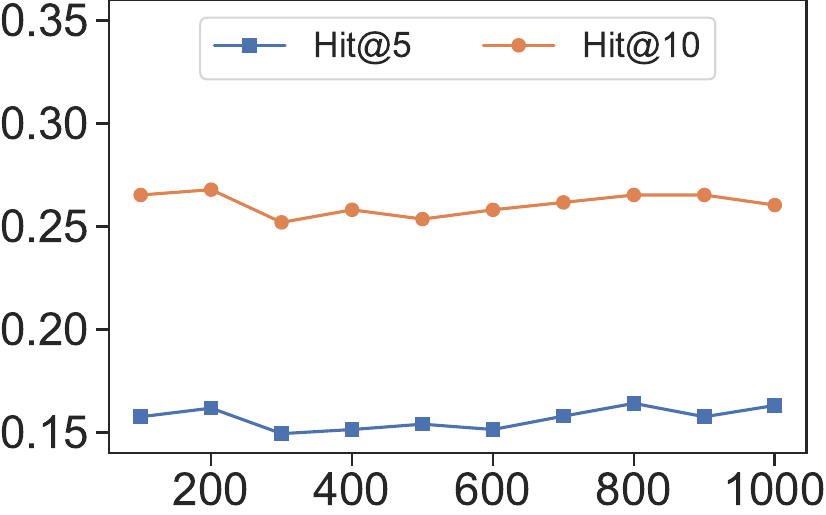}}
  \caption{ The influence of the number prototypes \wrt Hit.}
	\label{fig:prototypes2}
\end{figure}

\begin{figure}[h]
\vspace{-1.25em}
	\centering
	\subfigure[Travel]{\includegraphics[width=0.32\columnwidth]{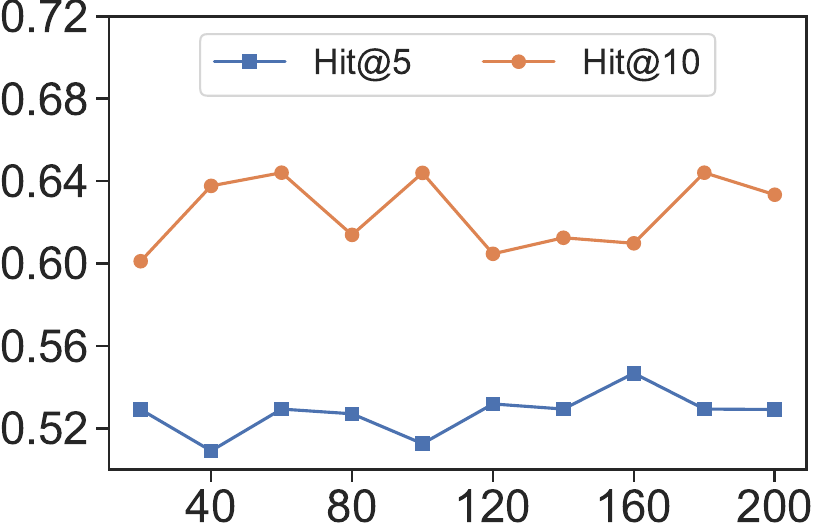}}
	\subfigure[Digital Art]{\includegraphics[width=0.32\columnwidth]{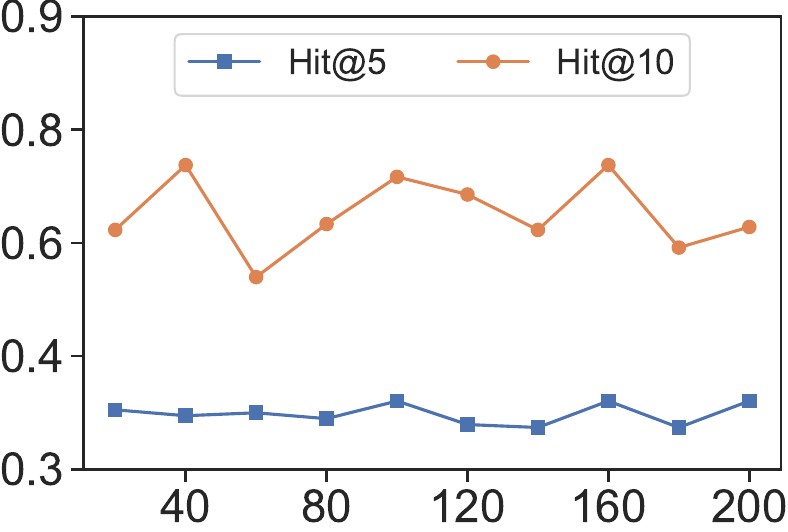}}
 	\subfigure[Rental]{\includegraphics[width=0.32\columnwidth]{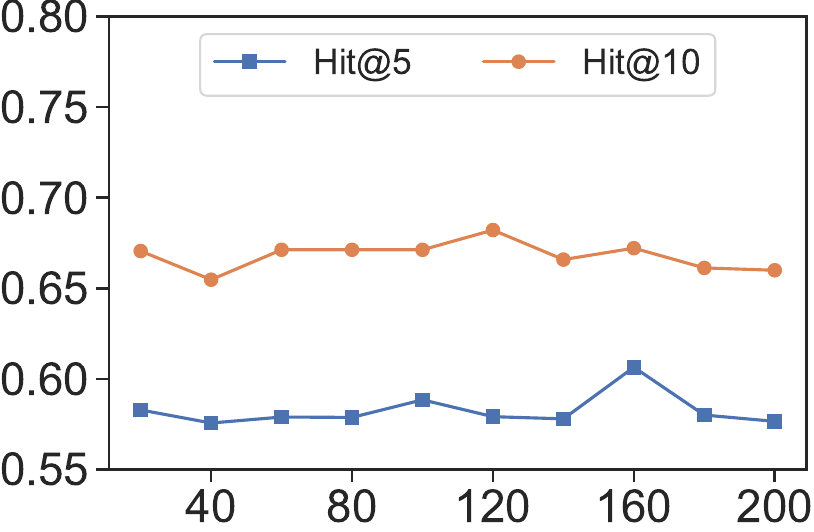}}
  
        \subfigure[Food Delivery]{\includegraphics[width=0.32\columnwidth]{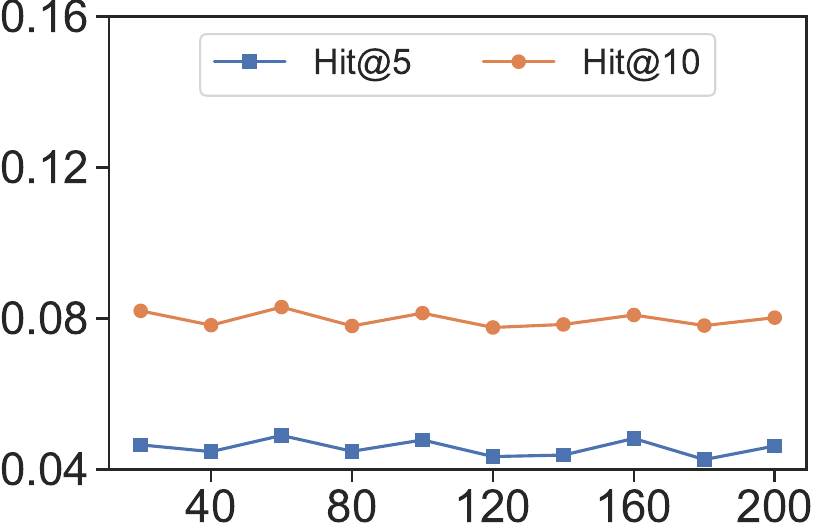}}
        \subfigure[Daily Necessity]{\includegraphics[width=0.32\columnwidth]{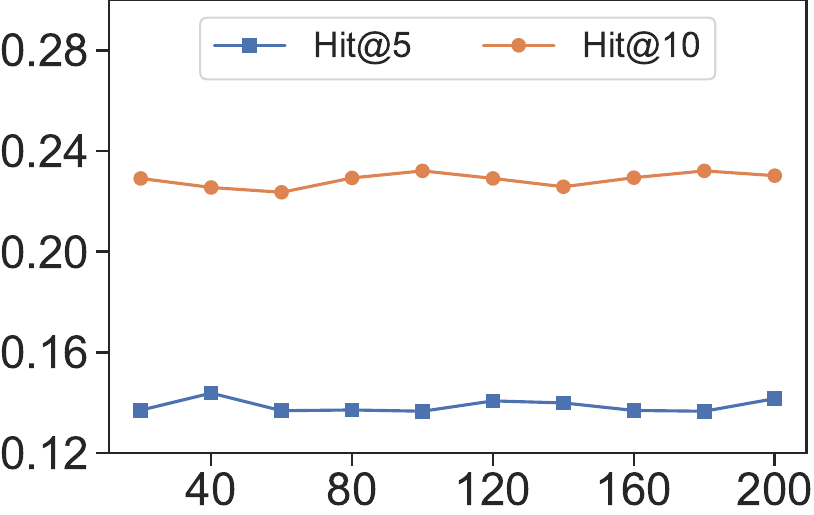}}
        \subfigure[Dining]{\includegraphics[width=0.32\columnwidth]{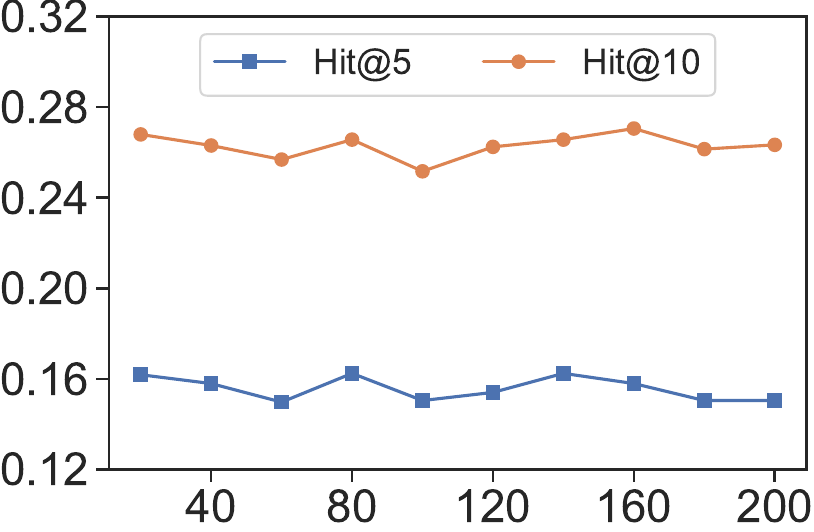}}
  \caption{ The study of the prototype enhanced attention \wrt Hit.}
	\label{fig:topk2}
	\vspace{-2em}
\end{figure}

 In the paper we only present performance comparison \wrt the NDCG metric for the in-depth anaslysis of {\model}. To comprehensively understand the merits of {\model}, we further provide additional comparison results \wrt the Hit metric. 
Specifically, we report the  ablation study in Figure~\ref{fig:ab_study2}, the influence of the number prototypes in Figure~\ref{fig:prototypes2}, and  the study of the prototype enhanced attention in Figure~\ref{fig:topk2}. Overall, compared to the results \wrt the NDCG metric, the similar trends and findings are concluded.

\subsubsection{Efficiency Analysis}
To illustrate the efficiency of {\model}, firstly we report the time cost (in minutes) of each method in the Pre-training stage as follows: SASRec (670), BERT4Rec (675), PeterRec (420), PAUP (577), ComiRec (400), SimGCL (618), PCRec-L (572), PCRec-D (607), and PEACE (545), from which the efficiency of {\model} in training stage is verified.
Secondly, we show the decrease of response time (RT) in online serving, which plays a key role in measuring the efficiency of 
industrial recommender systems. Benefiting from our lightweight deployment, the RT has substantially reduced from 70ms to 40ms by minimizing the time taken to infer representation for users and entities during model training. Overall, this approach has empowered us to handle a wide range of requests efficiently, thereby augmenting the scalability of the system.

%% file: main.bbl

\begin{thebibliography}{64}


\ifx \showCODEN    \undefined \def \showCODEN     #1{\unskip}     \fi
\ifx \showDOI      \undefined \def \showDOI       #1{#1}\fi
\ifx \showISBNx    \undefined \def \showISBNx     #1{\unskip}     \fi
\ifx \showISBNxiii \undefined \def \showISBNxiii  #1{\unskip}     \fi
\ifx \showISSN     \undefined \def \showISSN      #1{\unskip}     \fi
\ifx \showLCCN     \undefined \def \showLCCN      #1{\unskip}     \fi
\ifx \shownote     \undefined \def \shownote      #1{#1}          \fi
\ifx \showarticletitle \undefined \def \showarticletitle #1{#1}   \fi
\ifx \showURL      \undefined \def \showURL       {\relax}        \fi
\providecommand\bibfield[2]{#2}
\providecommand\bibinfo[2]{#2}
\providecommand\natexlab[1]{#1}
\providecommand\showeprint[2][]{arXiv:#2}

\bibitem[Aggarwal and Reddy(2014)]%
        {aggarwal2013}
\bibfield{author}{\bibinfo{person}{Charu~C. Aggarwal} {and}
  \bibinfo{person}{Chandan~K. Reddy}.} \bibinfo{year}{2014}\natexlab{}.
\newblock \showarticletitle{Data Clustering: Algorithms and Applications}. In
  \bibinfo{booktitle}{\emph{{CRC} Press}}.
\newblock


\bibitem[Arthur and Vassilvitskii(2007)]%
        {kmeans++}
\bibfield{author}{\bibinfo{person}{David Arthur} {and} \bibinfo{person}{Sergei
  Vassilvitskii}.} \bibinfo{year}{2007}\natexlab{}.
\newblock \showarticletitle{K-Means++: The Advantages of Careful Seeding}. In
  \bibinfo{booktitle}{\emph{SIAM}}. \bibinfo{pages}{1027–1035}.
\newblock


\bibitem[Cen et~al\mbox{.}(2020)]%
        {comirec2020}
\bibfield{author}{\bibinfo{person}{Yukuo Cen}, \bibinfo{person}{Jianwei Zhang},
  \bibinfo{person}{Xu Zou}, \bibinfo{person}{Chang Zhou},
  \bibinfo{person}{Hongxia Yang}, {and} \bibinfo{person}{Jie Tang}.}
  \bibinfo{year}{2020}\natexlab{}.
\newblock \showarticletitle{Controllable multi-interest framework for
  recommendation}. In \bibinfo{booktitle}{\emph{KDD}}.
  \bibinfo{pages}{2942--2951}.
\newblock


\bibitem[Choi et~al\mbox{.}(2023)]%
        {choi2023blurring}
\bibfield{author}{\bibinfo{person}{Jeongwhan Choi}, \bibinfo{person}{Seoyoung
  Hong}, \bibinfo{person}{Noseong Park}, {and} \bibinfo{person}{Sung-Bae Cho}.}
  \bibinfo{year}{2023}\natexlab{}.
\newblock \showarticletitle{Blurring-Sharpening Process Models for
  Collaborative Filtering}. In \bibinfo{booktitle}{\emph{SIGIR}}.
  \bibinfo{pages}{1096--1106}.
\newblock


\bibitem[der Maaten and Hinton(2008)]%
        {tsne2008}
\bibfield{author}{\bibinfo{person}{Laurens~Van der Maaten} {and}
  \bibinfo{person}{Geoffrey Hinton}.} \bibinfo{year}{2008}\natexlab{}.
\newblock \showarticletitle{Visualizing Data using t-SNE}. In
  \bibinfo{booktitle}{\emph{JMLR}}. \bibinfo{pages}{2579--2605}.
\newblock


\bibitem[Devlin et~al\mbox{.}(2019)]%
        {bert2019}
\bibfield{author}{\bibinfo{person}{Jacob Devlin}, \bibinfo{person}{Ming{-}Wei
  Chang}, \bibinfo{person}{Kenton Lee}, {and} \bibinfo{person}{Kristina
  Toutanova}.} \bibinfo{year}{2019}\natexlab{}.
\newblock \showarticletitle{{BERT:} Pre-training of Deep Bidirectional
  Transformers for Language Understanding}. In
  \bibinfo{booktitle}{\emph{NAACL}}. \bibinfo{pages}{4171--4186}.
\newblock


\bibitem[Fan et~al\mbox{.}(2019)]%
        {fan2019graph}
\bibfield{author}{\bibinfo{person}{Wenqi Fan}, \bibinfo{person}{Yao Ma},
  \bibinfo{person}{Qing Li}, \bibinfo{person}{Yuan He}, \bibinfo{person}{Eric
  Zhao}, \bibinfo{person}{Jiliang Tang}, {and} \bibinfo{person}{Dawei Yin}.}
  \bibinfo{year}{2019}\natexlab{}.
\newblock \showarticletitle{Graph neural networks for social recommendation}.
  In \bibinfo{booktitle}{\emph{WWW}}. \bibinfo{pages}{417--426}.
\newblock


\bibitem[Gao et~al\mbox{.}(2023)]%
        {gao2023leveraging}
\bibfield{author}{\bibinfo{person}{Weibo Gao}, \bibinfo{person}{Hao Wang},
  \bibinfo{person}{Qi Liu}, \bibinfo{person}{Fei Wang}, \bibinfo{person}{Xin
  Lin}, \bibinfo{person}{Linan Yue}, \bibinfo{person}{Zheng Zhang},
  \bibinfo{person}{Rui Lv}, {and} \bibinfo{person}{Shijin Wang}.}
  \bibinfo{year}{2023}\natexlab{}.
\newblock \showarticletitle{Leveraging Transferable Knowledge Concept Graph
  Embedding for Cold-Start Cognitive Diagnosis}. In
  \bibinfo{booktitle}{\emph{SIGIR}}. \bibinfo{pages}{983--992}.
\newblock


\bibitem[Gu et~al\mbox{.}(2021)]%
        {gu2021self}
\bibfield{author}{\bibinfo{person}{Yulong Gu}, \bibinfo{person}{Wentian Bao},
  \bibinfo{person}{Dan Ou}, \bibinfo{person}{Xiang Li},
  \bibinfo{person}{Baoliang Cui}, \bibinfo{person}{Biyu Ma},
  \bibinfo{person}{Haikuan Huang}, \bibinfo{person}{Qingwen Liu}, {and}
  \bibinfo{person}{Xiaoyi Zeng}.} \bibinfo{year}{2021}\natexlab{}.
\newblock \showarticletitle{Self-Supervised Learning on Users' Spontaneous
  Behaviors for Multi-Scenario Ranking in E-commerce}. In
  \bibinfo{booktitle}{\emph{CIKM}}. \bibinfo{pages}{3828--3837}.
\newblock


\bibitem[Guo et~al\mbox{.}(2017)]%
        {deepfm2017}
\bibfield{author}{\bibinfo{person}{Huifeng Guo}, \bibinfo{person}{Ruiming
  Tang}, \bibinfo{person}{Yunming Ye}, \bibinfo{person}{Zhenguo Li}, {and}
  \bibinfo{person}{Xiuqiang He}.} \bibinfo{year}{2017}\natexlab{}.
\newblock \showarticletitle{DeepFM: {A} Factorization-Machine based Neural
  Network for {CTR} Prediction}. In \bibinfo{booktitle}{\emph{IJCAI}}.
  \bibinfo{pages}{1725--1731}.
\newblock


\bibitem[He et~al\mbox{.}(2022)]%
        {mae2022}
\bibfield{author}{\bibinfo{person}{Kaiming He}, \bibinfo{person}{Xinlei Chen},
  \bibinfo{person}{Saining Xie}, \bibinfo{person}{Yanghao Li},
  \bibinfo{person}{Piotr Doll{\'{a}}r}, {and} \bibinfo{person}{Ross~B.
  Girshick}.} \bibinfo{year}{2022}\natexlab{}.
\newblock \showarticletitle{Masked Autoencoders Are Scalable Vision Learners}.
  In \bibinfo{booktitle}{\emph{CVPR}}. \bibinfo{pages}{15979--15988}.
\newblock


\bibitem[He et~al\mbox{.}(2020)]%
        {lightgcn2020}
\bibfield{author}{\bibinfo{person}{Xiangnan He}, \bibinfo{person}{Kuan Deng},
  \bibinfo{person}{Xiang Wang}, \bibinfo{person}{Yan Li},
  \bibinfo{person}{Yongdong Zhang}, {and} \bibinfo{person}{Meng Wang}.}
  \bibinfo{year}{2020}\natexlab{}.
\newblock \showarticletitle{Lightgcn: Simplifying and powering graph
  convolution network for recommendation}. In
  \bibinfo{booktitle}{\emph{SIGIR}}. \bibinfo{pages}{639--648}.
\newblock


\bibitem[He et~al\mbox{.}(2017)]%
        {neural2017he}
\bibfield{author}{\bibinfo{person}{Xiangnan He}, \bibinfo{person}{Lizi Liao},
  \bibinfo{person}{Hanwang Zhang}, \bibinfo{person}{Liqiang Nie},
  \bibinfo{person}{Xia Hu}, {and} \bibinfo{person}{Tat-Seng Chua}.}
  \bibinfo{year}{2017}\natexlab{}.
\newblock \showarticletitle{Neural Collaborative Filtering}. In
  \bibinfo{booktitle}{\emph{WWW}}. \bibinfo{pages}{173–182}.
\newblock


\bibitem[Hou et~al\mbox{.}(2022)]%
        {hou2022towards}
\bibfield{author}{\bibinfo{person}{Yupeng Hou}, \bibinfo{person}{Shanlei Mu},
  \bibinfo{person}{Wayne~Xin Zhao}, \bibinfo{person}{Yaliang Li},
  \bibinfo{person}{Bolin Ding}, {and} \bibinfo{person}{Ji-Rong Wen}.}
  \bibinfo{year}{2022}\natexlab{}.
\newblock \showarticletitle{Towards Universal Sequence Representation Learning
  for Recommender Systems}. In \bibinfo{booktitle}{\emph{KDD}}.
  \bibinfo{pages}{585--593}.
\newblock


\bibitem[Hu et~al\mbox{.}(2018)]%
        {hu2018leveraging}
\bibfield{author}{\bibinfo{person}{Binbin Hu}, \bibinfo{person}{Chuan Shi},
  \bibinfo{person}{Wayne~Xin Zhao}, {and} \bibinfo{person}{Philip~S Yu}.}
  \bibinfo{year}{2018}\natexlab{}.
\newblock \showarticletitle{Leveraging meta-path based context for top-n
  recommendation with a neural co-attention model}. In
  \bibinfo{booktitle}{\emph{KDD}}. \bibinfo{pages}{1531--1540}.
\newblock


\bibitem[Huai et~al\mbox{.}(2023)]%
        {huai2023m2gnn}
\bibfield{author}{\bibinfo{person}{Zepeng Huai}, \bibinfo{person}{Yuji Yang},
  \bibinfo{person}{Mengdi Zhang}, \bibinfo{person}{Zhongyi Zhang},
  \bibinfo{person}{Yichun Li}, {and} \bibinfo{person}{Wei Wu}.}
  \bibinfo{year}{2023}\natexlab{}.
\newblock \showarticletitle{M2GNN: Metapath and Multi-Interest Aggregated Graph
  Neural Network for Tag-Based Cross-Domain Recommendation}. In
  \bibinfo{booktitle}{\emph{SIGIR}}. \bibinfo{pages}{1468–1477}.
\newblock


\bibitem[Huang et~al\mbox{.}(2021)]%
        {huang2021knowledge}
\bibfield{author}{\bibinfo{person}{Chao Huang}, \bibinfo{person}{Huance Xu},
  \bibinfo{person}{Yong Xu}, \bibinfo{person}{Peng Dai},
  \bibinfo{person}{Lianghao Xia}, \bibinfo{person}{Mengyin Lu},
  \bibinfo{person}{Liefeng Bo}, \bibinfo{person}{Hao Xing},
  \bibinfo{person}{Xiaoping Lai}, {and} \bibinfo{person}{Yanfang Ye}.}
  \bibinfo{year}{2021}\natexlab{}.
\newblock \showarticletitle{Knowledge-aware coupled graph neural network for
  social recommendation}. In \bibinfo{booktitle}{\emph{AAAI}}.
  \bibinfo{pages}{4115--4122}.
\newblock


\bibitem[Kang et~al\mbox{.}(2019)]%
        {semitrans2019}
\bibfield{author}{\bibinfo{person}{SeongKu Kang}, \bibinfo{person}{Junyoung
  Hwang}, \bibinfo{person}{Dongha Lee}, {and} \bibinfo{person}{Hwanjo Yu}.}
  \bibinfo{year}{2019}\natexlab{}.
\newblock \showarticletitle{Semi-Supervised Learning for Cross-Domain
  Recommendation to Cold-Start Users}. In \bibinfo{booktitle}{\emph{CIKM}}.
  \bibinfo{pages}{1563–1572}.
\newblock


\bibitem[Kang and McAuley(2018)]%
        {sasrec2018}
\bibfield{author}{\bibinfo{person}{Wang-Cheng Kang} {and}
  \bibinfo{person}{Julian McAuley}.} \bibinfo{year}{2018}\natexlab{}.
\newblock \showarticletitle{Self-attentive sequential recommendation}. In
  \bibinfo{booktitle}{\emph{ICDM}}. \bibinfo{pages}{197--206}.
\newblock


\bibitem[Krishnan et~al\mbox{.}(2020)]%
        {MMT-Net2020}
\bibfield{author}{\bibinfo{person}{Adit Krishnan}, \bibinfo{person}{Mahashweta
  Das}, \bibinfo{person}{Mangesh Bendre}, \bibinfo{person}{Hao Yang}, {and}
  \bibinfo{person}{Hari Sundaram}.} \bibinfo{year}{2020}\natexlab{}.
\newblock \showarticletitle{Transfer Learning via Contextual Invariants for
  One-to-Many Cross-Domain Recommendation}. In
  \bibinfo{booktitle}{\emph{SIGIR}}. \bibinfo{pages}{1081–1090}.
\newblock


\bibitem[Lample et~al\mbox{.}(2016)]%
        {lample2016ner}
\bibfield{author}{\bibinfo{person}{Guillaume Lample}, \bibinfo{person}{Miguel
  Ballesteros}, \bibinfo{person}{Sandeep Subramanian}, \bibinfo{person}{Kazuya
  Kawakami}, {and} \bibinfo{person}{Chris Dyer}.}
  \bibinfo{year}{2016}\natexlab{}.
\newblock \showarticletitle{Neural Architectures for Named Entity Recognition}.
  In \bibinfo{booktitle}{\emph{NAACL}}. \bibinfo{pages}{260--270}.
\newblock


\bibitem[Li et~al\mbox{.}(2022)]%
        {disengnn2022}
\bibfield{author}{\bibinfo{person}{Ansong Li}, \bibinfo{person}{Zhiyong Cheng},
  \bibinfo{person}{Fan Liu}, \bibinfo{person}{Zan Gao}, \bibinfo{person}{Weili
  Guan}, {and} \bibinfo{person}{Yuxin Peng}.} \bibinfo{year}{2022}\natexlab{}.
\newblock \showarticletitle{Disentangled Graph Neural Networks for
  Session-based Recommendation}. In \bibinfo{booktitle}{\emph{TKDE}}.
\newblock


\bibitem[Li et~al\mbox{.}(2023)]%
        {li2023one}
\bibfield{author}{\bibinfo{person}{Chenglin Li}, \bibinfo{person}{Yuanzhen
  Xie}, \bibinfo{person}{Chenyun Yu}, \bibinfo{person}{Bo Hu},
  \bibinfo{person}{Zang Li}, \bibinfo{person}{Guoqiang Shu},
  \bibinfo{person}{Xiaohu Qie}, {and} \bibinfo{person}{Di Niu}.}
  \bibinfo{year}{2023}\natexlab{}.
\newblock \showarticletitle{One for All, All for One: Learning and Transferring
  User Embeddings for Cross-Domain Recommendation}. In
  \bibinfo{booktitle}{\emph{WSDM}}. \bibinfo{pages}{366--374}.
\newblock


\bibitem[Li and Tuzhilin(2020)]%
        {ddtcdr2020}
\bibfield{author}{\bibinfo{person}{Pan Li} {and} \bibinfo{person}{Alexander
  Tuzhilin}.} \bibinfo{year}{2020}\natexlab{}.
\newblock \showarticletitle{{DDTCDR:} Deep Dual Transfer Cross Domain
  Recommendation}. In \bibinfo{booktitle}{\emph{WSDM}}.
  \bibinfo{pages}{331--339}.
\newblock


\bibitem[Liu et~al\mbox{.}(2020)]%
        {bitrans2020}
\bibfield{author}{\bibinfo{person}{Meng Liu}, \bibinfo{person}{Jianjun Li},
  \bibinfo{person}{Guohui Li}, {and} \bibinfo{person}{Peng Pan}.}
  \bibinfo{year}{2020}\natexlab{}.
\newblock \showarticletitle{Cross Domain Recommendation via Bi-Directional
  Transfer Graph Collaborative Filtering Networks}. In
  \bibinfo{booktitle}{\emph{CIKM}}. \bibinfo{pages}{885–894}.
\newblock


\bibitem[Liu et~al\mbox{.}(2023)]%
        {liu2023pre}
\bibfield{author}{\bibinfo{person}{Peng Liu}, \bibinfo{person}{Lemei Zhang},
  {and} \bibinfo{person}{Jon~Atle Gulla}.} \bibinfo{year}{2023}\natexlab{}.
\newblock \showarticletitle{Pre-train, prompt and recommendation: A
  comprehensive survey of language modelling paradigm adaptations in
  recommender systems}. In \bibinfo{booktitle}{\emph{arXiv preprint
  arXiv:2302.03735}}.
\newblock


\bibitem[Lu et~al\mbox{.}(2023)]%
        {lu2023three}
\bibfield{author}{\bibinfo{person}{Jinhu Lu}, \bibinfo{person}{Guohao Sun},
  \bibinfo{person}{Xiu Fang}, \bibinfo{person}{Jian Yang}, {and}
  \bibinfo{person}{Wei He}.} \bibinfo{year}{2023}\natexlab{}.
\newblock \showarticletitle{A Three-Layer Attentional Framework Based on
  Similar Users for Dual-Target Cross-Domain Recommendation}. In
  \bibinfo{booktitle}{\emph{DASFAA}}. \bibinfo{pages}{297--313}.
\newblock


\bibitem[Lu et~al\mbox{.}(2020)]%
        {lu2020meta}
\bibfield{author}{\bibinfo{person}{Yuanfu Lu}, \bibinfo{person}{Yuan Fang},
  {and} \bibinfo{person}{Chuan Shi}.} \bibinfo{year}{2020}\natexlab{}.
\newblock \showarticletitle{Meta-learning on heterogeneous information networks
  for cold-start recommendation}. In \bibinfo{booktitle}{\emph{KDD}}.
  \bibinfo{pages}{1563--1573}.
\newblock


\bibitem[Ma et~al\mbox{.}(2020)]%
        {ma2020disentangled}
\bibfield{author}{\bibinfo{person}{Jianxin Ma}, \bibinfo{person}{Chang Zhou},
  \bibinfo{person}{Hongxia Yang}, \bibinfo{person}{Peng Cui},
  \bibinfo{person}{Xin Wang}, {and} \bibinfo{person}{Wenwu Zhu}.}
  \bibinfo{year}{2020}\natexlab{}.
\newblock \showarticletitle{Disentangled self-supervision in sequential
  recommenders}. In \bibinfo{booktitle}{\emph{KDD}}. \bibinfo{pages}{483--491}.
\newblock


\bibitem[Ma et~al\mbox{.}(2018)]%
        {esmm}
\bibfield{author}{\bibinfo{person}{Xiao Ma}, \bibinfo{person}{Liqin Zhao},
  \bibinfo{person}{Guan Huang}, \bibinfo{person}{Zhi Wang},
  \bibinfo{person}{Zelin Hu}, \bibinfo{person}{Xiaoqiang Zhu}, {and}
  \bibinfo{person}{Kun Gai}.} \bibinfo{year}{2018}\natexlab{}.
\newblock \showarticletitle{Entire Space Multi-Task Model: An Effective
  Approach for Estimating Post-Click Conversion Rate}. In
  \bibinfo{booktitle}{\emph{SIGIR}}. \bibinfo{pages}{1137–1140}.
\newblock


\bibitem[Man et~al\mbox{.}(2017)]%
        {emcdr2017}
\bibfield{author}{\bibinfo{person}{Tong Man}, \bibinfo{person}{Huawei Shen},
  \bibinfo{person}{Xiaolong Jin}, {and} \bibinfo{person}{Xueqi Cheng}.}
  \bibinfo{year}{2017}\natexlab{}.
\newblock \showarticletitle{Cross-Domain Recommendation: An Embedding and
  Mapping Approach}. In \bibinfo{booktitle}{\emph{IJCAI}}.
  \bibinfo{pages}{2464–2470}.
\newblock


\bibitem[Pi et~al\mbox{.}(2019)]%
        {PiBZZG19}
\bibfield{author}{\bibinfo{person}{Qi Pi}, \bibinfo{person}{Weijie Bian},
  \bibinfo{person}{Guorui Zhou}, \bibinfo{person}{Xiaoqiang Zhu}, {and}
  \bibinfo{person}{Kun Gai}.} \bibinfo{year}{2019}\natexlab{}.
\newblock \showarticletitle{Practice on Long Sequential User Behavior Modeling
  for Click-Through Rate Prediction}. In \bibinfo{booktitle}{\emph{KDD}}.
  \bibinfo{pages}{2671--2679}.
\newblock


\bibitem[Sun et~al\mbox{.}(2019)]%
        {bert4rec2019}
\bibfield{author}{\bibinfo{person}{Fei Sun}, \bibinfo{person}{Jun Liu},
  \bibinfo{person}{Jian Wu}, \bibinfo{person}{Changhua Pei},
  \bibinfo{person}{Xiao Lin}, \bibinfo{person}{Wenwu Ou}, {and}
  \bibinfo{person}{Peng Jiang}.} \bibinfo{year}{2019}\natexlab{}.
\newblock \showarticletitle{BERT4Rec: Sequential Recommendation with
  Bidirectional Encoder Representations from Transformer}. In
  \bibinfo{booktitle}{\emph{CIKM}}. \bibinfo{pages}{1441–1450}.
\newblock


\bibitem[van~den Oord et~al\mbox{.}(2018)]%
        {infonce2018}
\bibfield{author}{\bibinfo{person}{A{\"{a}}ron van~den Oord},
  \bibinfo{person}{Yazhe Li}, {and} \bibinfo{person}{Oriol Vinyals}.}
  \bibinfo{year}{2018}\natexlab{}.
\newblock \showarticletitle{Representation Learning with Contrastive Predictive
  Coding}. In \bibinfo{booktitle}{\emph{arXiv preprint arXiv:1807.03748}}.
\newblock


\bibitem[Vaswani et~al\mbox{.}(2017)]%
        {transformer2017}
\bibfield{author}{\bibinfo{person}{Ashish Vaswani}, \bibinfo{person}{Noam
  Shazeer}, \bibinfo{person}{Niki Parmar}, \bibinfo{person}{Jakob Uszkoreit},
  \bibinfo{person}{Llion Jones}, \bibinfo{person}{Aidan~N. Gomez},
  \bibinfo{person}{Lukasz Kaiser}, {and} \bibinfo{person}{Illia Polosukhin}.}
  \bibinfo{year}{2017}\natexlab{}.
\newblock \showarticletitle{Attention is All you Need}. In
  \bibinfo{booktitle}{\emph{NIPS}}. \bibinfo{pages}{5998--6008}.
\newblock


\bibitem[Velickovic et~al\mbox{.}(2018)]%
        {gat2018}
\bibfield{author}{\bibinfo{person}{Petar Velickovic}, \bibinfo{person}{Guillem
  Cucurull}, \bibinfo{person}{Arantxa Casanova}, \bibinfo{person}{Adriana
  Romero}, \bibinfo{person}{Pietro Li{\`{o}}}, {and} \bibinfo{person}{Yoshua
  Bengio}.} \bibinfo{year}{2018}\natexlab{}.
\newblock \showarticletitle{Graph Attention Networks}. In
  \bibinfo{booktitle}{\emph{ICLR}}.
\newblock


\bibitem[Wang et~al\mbox{.}(2021b)]%
        {wang2021pre}
\bibfield{author}{\bibinfo{person}{Chen Wang}, \bibinfo{person}{Yueqing Liang},
  \bibinfo{person}{Zhiwei Liu}, \bibinfo{person}{Tao Zhang}, {and}
  \bibinfo{person}{S~Yu Philip}.} \bibinfo{year}{2021}\natexlab{b}.
\newblock \showarticletitle{Pre-training graph neural network for cross domain
  recommendation}. In \bibinfo{booktitle}{\emph{CogMI}}.
  \bibinfo{pages}{140--145}.
\newblock


\bibitem[Wang and Liu(2021)]%
        {wang2021under}
\bibfield{author}{\bibinfo{person}{Feng Wang} {and} \bibinfo{person}{Huaping
  Liu}.} \bibinfo{year}{2021}\natexlab{}.
\newblock \showarticletitle{Understanding the Behaviour of Contrastive Loss}.
  In \bibinfo{booktitle}{\emph{CVPR}}. \bibinfo{pages}{2495--2504}.
\newblock


\bibitem[Wang and Lv(2020)]%
        {commonfeatures2020}
\bibfield{author}{\bibinfo{person}{Jiaqi Wang} {and} \bibinfo{person}{Jing
  Lv}.} \bibinfo{year}{2020}\natexlab{}.
\newblock \showarticletitle{Tag-informed collaborative topic modeling for cross
  domain recommendations}. In \bibinfo{booktitle}{\emph{KBS}}.
  \bibinfo{pages}{106--119}.
\newblock


\bibitem[Wang et~al\mbox{.}(2019a)]%
        {wang2019kgat}
\bibfield{author}{\bibinfo{person}{Xiang Wang}, \bibinfo{person}{Xiangnan He},
  \bibinfo{person}{Yixin Cao}, \bibinfo{person}{Meng Liu}, {and}
  \bibinfo{person}{Tat-Seng Chua}.} \bibinfo{year}{2019}\natexlab{a}.
\newblock \showarticletitle{Kgat: Knowledge graph attention network for
  recommendation}. In \bibinfo{booktitle}{\emph{KDD}}.
  \bibinfo{pages}{950--958}.
\newblock


\bibitem[Wang et~al\mbox{.}(2019b)]%
        {wang2019neural}
\bibfield{author}{\bibinfo{person}{Xiang Wang}, \bibinfo{person}{Xiangnan He},
  \bibinfo{person}{Meng Wang}, \bibinfo{person}{Fuli Feng}, {and}
  \bibinfo{person}{Tat-Seng Chua}.} \bibinfo{year}{2019}\natexlab{b}.
\newblock \showarticletitle{Neural graph collaborative filtering}. In
  \bibinfo{booktitle}{\emph{SIGIR}}. \bibinfo{pages}{165--174}.
\newblock


\bibitem[Wang et~al\mbox{.}(2021a)]%
        {wang2021learning}
\bibfield{author}{\bibinfo{person}{Xiang Wang}, \bibinfo{person}{Tinglin
  Huang}, \bibinfo{person}{Dingxian Wang}, \bibinfo{person}{Yancheng Yuan},
  \bibinfo{person}{Zhenguang Liu}, \bibinfo{person}{Xiangnan He}, {and}
  \bibinfo{person}{Tat-Seng Chua}.} \bibinfo{year}{2021}\natexlab{a}.
\newblock \showarticletitle{Learning intents behind interactions with knowledge
  graph for recommendation}. In \bibinfo{booktitle}{\emph{WWW}}.
  \bibinfo{pages}{878--887}.
\newblock


\bibitem[Wang et~al\mbox{.}(2019c)]%
        {wang2019explainable}
\bibfield{author}{\bibinfo{person}{Xiang Wang}, \bibinfo{person}{Dingxian
  Wang}, \bibinfo{person}{Canran Xu}, \bibinfo{person}{Xiangnan He},
  \bibinfo{person}{Yixin Cao}, {and} \bibinfo{person}{Tat-Seng Chua}.}
  \bibinfo{year}{2019}\natexlab{c}.
\newblock \showarticletitle{Explainable reasoning over knowledge graphs for
  recommendation}. In \bibinfo{booktitle}{\emph{AAAI}}.
  \bibinfo{pages}{5329--5336}.
\newblock


\bibitem[Wang et~al\mbox{.}(2018)]%
        {negativetransfer2019}
\bibfield{author}{\bibinfo{person}{Zirui Wang}, \bibinfo{person}{Zihang Dai},
  \bibinfo{person}{Barnabás Póczos}, {and} \bibinfo{person}{Jaime~G.
  Carbonell}.} \bibinfo{year}{2018}\natexlab{}.
\newblock \showarticletitle{Characterizing and Avoiding Negative Transfer}. In
  \bibinfo{booktitle}{\emph{CVPR}}. \bibinfo{pages}{11293--11302}.
\newblock


\bibitem[Wei et~al\mbox{.}(2023)]%
        {wei2023lightgt}
\bibfield{author}{\bibinfo{person}{Yinwei Wei}, \bibinfo{person}{Wenqi Liu},
  \bibinfo{person}{Fan Liu}, \bibinfo{person}{Xiang Wang},
  \bibinfo{person}{Liqiang Nie}, {and} \bibinfo{person}{Tat-Seng Chua}.}
  \bibinfo{year}{2023}\natexlab{}.
\newblock \showarticletitle{LightGT: A Light Graph Transformer for Multimedia
  Recommendation}. In \bibinfo{booktitle}{\emph{SIGIR}}.
  \bibinfo{pages}{1508--1517}.
\newblock


\bibitem[Wen et~al\mbox{.}(2020)]%
        {esm22020}
\bibfield{author}{\bibinfo{person}{Hong Wen}, \bibinfo{person}{Jing Zhang},
  \bibinfo{person}{Yuan Wang}, \bibinfo{person}{Fuyu Lv},
  \bibinfo{person}{Wentian Bao}, \bibinfo{person}{Quan Lin}, {and}
  \bibinfo{person}{Keping Yang}.} \bibinfo{year}{2020}\natexlab{}.
\newblock \showarticletitle{Entire Space Multi-Task Modeling via Post-Click
  Behavior Decomposition for Conversion Rate Prediction}. In
  \bibinfo{booktitle}{\emph{SIGIR}}. \bibinfo{pages}{2377–2386}.
\newblock


\bibitem[Wu et~al\mbox{.}(2021)]%
        {wu2021self}
\bibfield{author}{\bibinfo{person}{Jiancan Wu}, \bibinfo{person}{Xiang Wang},
  \bibinfo{person}{Fuli Feng}, \bibinfo{person}{Xiangnan He},
  \bibinfo{person}{Liang Chen}, \bibinfo{person}{Jianxun Lian}, {and}
  \bibinfo{person}{Xing Xie}.} \bibinfo{year}{2021}\natexlab{}.
\newblock \showarticletitle{Self-supervised graph learning for recommendation}.
  In \bibinfo{booktitle}{\emph{SIGIR}}. \bibinfo{pages}{726–735}.
\newblock


\bibitem[Wu et~al\mbox{.}(2022)]%
        {wu2022graph}
\bibfield{author}{\bibinfo{person}{Shiwen Wu}, \bibinfo{person}{Fei Sun},
  \bibinfo{person}{Wentao Zhang}, \bibinfo{person}{Xu Xie}, {and}
  \bibinfo{person}{Bin Cui}.} \bibinfo{year}{2022}\natexlab{}.
\newblock \showarticletitle{Graph neural networks in recommender systems: a
  survey}. In \bibinfo{booktitle}{\emph{CSUR}}. \bibinfo{pages}{1--37}.
\newblock


\bibitem[Wu et~al\mbox{.}(2020)]%
        {wu2020comprehensive}
\bibfield{author}{\bibinfo{person}{Zonghan Wu}, \bibinfo{person}{Shirui Pan},
  \bibinfo{person}{Fengwen Chen}, \bibinfo{person}{Guodong Long},
  \bibinfo{person}{Chengqi Zhang}, {and} \bibinfo{person}{S~Yu Philip}.}
  \bibinfo{year}{2020}\natexlab{}.
\newblock \showarticletitle{A comprehensive survey on graph neural networks}.
  In \bibinfo{booktitle}{\emph{TNNLS}}. \bibinfo{pages}{4--24}.
\newblock


\bibitem[Yao et~al\mbox{.}(2021)]%
        {yao2021self}
\bibfield{author}{\bibinfo{person}{Tiansheng Yao}, \bibinfo{person}{Xinyang
  Yi}, \bibinfo{person}{Derek~Zhiyuan Cheng}, \bibinfo{person}{Felix Yu},
  \bibinfo{person}{Ting Chen}, \bibinfo{person}{Aditya Menon},
  \bibinfo{person}{Lichan Hong}, \bibinfo{person}{Ed~H Chi},
  \bibinfo{person}{Steve Tjoa}, \bibinfo{person}{Jieqi Kang}, {et~al\mbox{.}}}
  \bibinfo{year}{2021}\natexlab{}.
\newblock \showarticletitle{Self-supervised learning for large-scale item
  recommendations}. In \bibinfo{booktitle}{\emph{CIKM}}.
  \bibinfo{pages}{4321--4330}.
\newblock


\bibitem[Yu et~al\mbox{.}(2021)]%
        {yu2021self}
\bibfield{author}{\bibinfo{person}{Junliang Yu}, \bibinfo{person}{Hongzhi Yin},
  \bibinfo{person}{Jundong Li}, \bibinfo{person}{Qinyong Wang},
  \bibinfo{person}{Nguyen Quoc~Viet Hung}, {and} \bibinfo{person}{Xiangliang
  Zhang}.} \bibinfo{year}{2021}\natexlab{}.
\newblock \showarticletitle{Self-supervised multi-channel hypergraph
  convolutional network for social recommendation}. In
  \bibinfo{booktitle}{\emph{WWW}}. \bibinfo{pages}{413--424}.
\newblock


\bibitem[Yu et~al\mbox{.}(2022)]%
        {simgcl2022}
\bibfield{author}{\bibinfo{person}{Junliang Yu}, \bibinfo{person}{Hongzhi Yin},
  \bibinfo{person}{Xin Xia}, \bibinfo{person}{Tong Chen},
  \bibinfo{person}{Lizhen Cui}, {and} \bibinfo{person}{Quoc Viet~Hung Nguyen}.}
  \bibinfo{year}{2022}\natexlab{}.
\newblock \showarticletitle{Are Graph Augmentations Necessary? Simple Graph
  Contrastive Learning for Recommendation}. In
  \bibinfo{booktitle}{\emph{SIGIR}}. \bibinfo{pages}{1294–1303}.
\newblock


\bibitem[Yuan et~al\mbox{.}(2020)]%
        {peterrec2020}
\bibfield{author}{\bibinfo{person}{Fajie Yuan}, \bibinfo{person}{Xiangnan He},
  \bibinfo{person}{Alexandros Karatzoglou}, {and} \bibinfo{person}{Liguang
  Zhang}.} \bibinfo{year}{2020}\natexlab{}.
\newblock \showarticletitle{Parameter-Efficient Transfer from Sequential
  Behaviors for User Modeling and Recommendation}. In
  \bibinfo{booktitle}{\emph{SIGIR}}. \bibinfo{pages}{1469--1478}.
\newblock


\bibitem[Zang et~al\mbox{.}(2022)]%
        {cdrsurvey2022}
\bibfield{author}{\bibinfo{person}{Tianzi Zang}, \bibinfo{person}{Yanmin Zhu},
  \bibinfo{person}{Haobing Liu}, \bibinfo{person}{Ruohan Zhang}, {and}
  \bibinfo{person}{Jiadi Yu}.} \bibinfo{year}{2022}\natexlab{}.
\newblock \showarticletitle{A Survey on Cross-Domain Recommendation:
  Taxonomies, Methods, and Future Directions}. In
  \bibinfo{booktitle}{\emph{TOIS}}. \bibinfo{pages}{1--39}.
\newblock


\bibitem[Zang et~al\mbox{.}(2023)]%
        {kddyiru}
\bibfield{author}{\bibinfo{person}{Xiaoling Zang}, \bibinfo{person}{Binbin Hu},
  \bibinfo{person}{Jun Chu}, \bibinfo{person}{Guannan Zhang},
  \bibinfo{person}{Zhiqiang Zhang}, \bibinfo{person}{Jun Zhou}, {and}
  \bibinfo{person}{Wenliang Zhong}.} \bibinfo{year}{2023}\natexlab{}.
\newblock \showarticletitle{Commonsense Knowledge Graph towards Super APP and
  Its Applications in Alipay}. In \bibinfo{booktitle}{\emph{KDD}}.
  \bibinfo{pages}{5509–5519}.
\newblock


\bibitem[Zhao et~al\mbox{.}(2022)]%
        {zhao2022multi}
\bibfield{author}{\bibinfo{person}{Xiaoyun Zhao}, \bibinfo{person}{Ning Yang},
  {and} \bibinfo{person}{Philip~S Yu}.} \bibinfo{year}{2022}\natexlab{}.
\newblock \showarticletitle{Multi-sparse-domain collaborative recommendation
  via enhanced comprehensive aspect preference learning}. In
  \bibinfo{booktitle}{\emph{WSDM}}. \bibinfo{pages}{1452--1460}.
\newblock


\bibitem[Zhou et~al\mbox{.}(2021)]%
        {zhou2021contrastive}
\bibfield{author}{\bibinfo{person}{Chang Zhou}, \bibinfo{person}{Jianxin Ma},
  \bibinfo{person}{Jianwei Zhang}, \bibinfo{person}{Jingren Zhou}, {and}
  \bibinfo{person}{Hongxia Yang}.} \bibinfo{year}{2021}\natexlab{}.
\newblock \showarticletitle{Contrastive learning for debiased candidate
  generation in large-scale recommender systems}. In
  \bibinfo{booktitle}{\emph{KDD}}. \bibinfo{pages}{3985--3995}.
\newblock


\bibitem[Zhu et~al\mbox{.}(2019a)]%
        {dtcdr2019}
\bibfield{author}{\bibinfo{person}{Feng Zhu}, \bibinfo{person}{Chaochao Chen},
  \bibinfo{person}{Yan Wang}, \bibinfo{person}{Guanfeng Liu}, {and}
  \bibinfo{person}{Xiaolin Zheng}.} \bibinfo{year}{2019}\natexlab{a}.
\newblock \showarticletitle{DTCDR: A Framework for Dual-Target Cross-Domain
  Recommendation}. In \bibinfo{booktitle}{\emph{CIKM}}.
  \bibinfo{pages}{1533–1542}.
\newblock


\bibitem[Zhu et~al\mbox{.}(2021b)]%
        {cdrsurvey2021}
\bibfield{author}{\bibinfo{person}{Feng Zhu}, \bibinfo{person}{Yan Wang},
  \bibinfo{person}{Chaochao Chen}, \bibinfo{person}{Jun Zhou},
  \bibinfo{person}{Longfei Li}, {and} \bibinfo{person}{Guanfeng Liu}.}
  \bibinfo{year}{2021}\natexlab{b}.
\newblock \showarticletitle{Cross-Domain Recommendation: Challenges, Progress,
  and Prospects}. In \bibinfo{booktitle}{\emph{IJCAI}}.
  \bibinfo{pages}{4721--4728}.
\newblock


\bibitem[Zhu et~al\mbox{.}(2019b)]%
        {Zhu2019re}
\bibfield{author}{\bibinfo{person}{Hao Zhu}, \bibinfo{person}{Yankai Lin},
  \bibinfo{person}{Zhiyuan Liu}, \bibinfo{person}{Jie Fu},
  \bibinfo{person}{Tat{-}Seng Chua}, {and} \bibinfo{person}{Maosong Sun}.}
  \bibinfo{year}{2019}\natexlab{b}.
\newblock \showarticletitle{Graph Neural Networks with Generated Parameters for
  Relation Extraction}. In \bibinfo{booktitle}{\emph{ACL}}.
  \bibinfo{pages}{1331--1339}.
\newblock


\bibitem[Zhu et~al\mbox{.}(2021a)]%
        {tmcdr2021}
\bibfield{author}{\bibinfo{person}{Yongchun Zhu}, \bibinfo{person}{Kaikai Ge},
  \bibinfo{person}{Fuzhen Zhuang}, \bibinfo{person}{Ruobing Xie},
  \bibinfo{person}{Dongbo Xi}, \bibinfo{person}{Xu Zhang},
  \bibinfo{person}{Leyu Lin}, {and} \bibinfo{person}{Qing He}.}
  \bibinfo{year}{2021}\natexlab{a}.
\newblock \showarticletitle{Transfer-Meta Framework for Cross-Domain
  Recommendation to Cold-Start Users}. In \bibinfo{booktitle}{\emph{SIGIR}}.
  \bibinfo{pages}{1813–1817}.
\newblock


\bibitem[Zhu et~al\mbox{.}(2022a)]%
        {paup2022}
\bibfield{author}{\bibinfo{person}{Yuehua Zhu}, \bibinfo{person}{Bo Huang},
  \bibinfo{person}{Shaohua Jiang}, \bibinfo{person}{Muli Yang},
  \bibinfo{person}{Yanhua Yang}, {and} \bibinfo{person}{Wenliang Zhong}.}
  \bibinfo{year}{2022}\natexlab{a}.
\newblock \showarticletitle{Progressive Self-Attention Network with
  Unsymmetrical Positional Encoding for Sequential Recommendation}. In
  \bibinfo{booktitle}{\emph{SIGIR}}. \bibinfo{pages}{2029–2033}.
\newblock


\bibitem[Zhu et~al\mbox{.}(2022b)]%
        {persontrans2022}
\bibfield{author}{\bibinfo{person}{Yongchun Zhu}, \bibinfo{person}{Zhenwei
  Tang}, \bibinfo{person}{Yudan Liu}, \bibinfo{person}{Fuzhen Zhuang},
  \bibinfo{person}{Ruobing Xie}, \bibinfo{person}{Xu Zhang},
  \bibinfo{person}{Leyu Lin}, {and} \bibinfo{person}{Qing He}.}
  \bibinfo{year}{2022}\natexlab{b}.
\newblock \showarticletitle{Personalized Transfer of User Preferences for
  Cross-domain Recommendation}. In \bibinfo{booktitle}{\emph{WSDM}}.
  \bibinfo{pages}{1507--1515}.
\newblock


\bibitem[Zhuo et~al\mbox{.}(2022)]%
        {tiger2022}
\bibfield{author}{\bibinfo{person}{Jianhuan Zhuo}, \bibinfo{person}{Jianxun
  Lian}, \bibinfo{person}{Lanling Xu}, \bibinfo{person}{Ming Gong},
  \bibinfo{person}{Linjun Shou}, \bibinfo{person}{Daxin Jiang},
  \bibinfo{person}{Xing Xie}, {and} \bibinfo{person}{Yinliang Yue}.}
  \bibinfo{year}{2022}\natexlab{}.
\newblock \showarticletitle{Tiger: Transferable Interest Graph Embedding for
  Domain-Level Zero-Shot Recommendation}. In \bibinfo{booktitle}{\emph{CIKM}}.
  \bibinfo{pages}{2806--2816}.
\newblock


\end{thebibliography}
